%% file: aanda.tex
\documentclass{aa}  

\usepackage{txfonts}

\newcommand{\logg} {\log \textsl{\textrm{g}}}

\newcommand{\Te} {T_{\rm eff} }

\newcommand{\msun} {\;$M_\odot$}
\newcommand{\mbol} {M_{\rm bol}}
\newcommand\gta{\lower 0.5ex\hbox{$\buildrel > \over \sim\ $}} 
\newcommand\lta{\lower 0.5ex\hbox{$\buildrel < \over \sim\ $}} 

\newcommand{\lsun} {L_{\odot}}

\newcommand{\Gbp}{G_\mathrm{BP}}
\newcommand{\Grp}{G_\mathrm{RP}}

\begin{document}

   \title{Classification and parameterisation of a large Gaia sample of white dwarfs using XP spectra}

   \author{O. Vincent
          \inst{1}, M.A. Barstow\inst{2}, S. Jordan\inst{3}, C. Mander\inst{2}, P. Bergeron\inst{1} \and \ P. Dufour\inst{1}
          }

   \institute{D\' epartement de Physique, Universit\' e\ de Montr\' eal, Montr\' eal, Qu\' ebec, Canada\\
              \email{o.vincent@umontreal.ca} \and School of Physics \& \ Astronomy, University of Leicester, Leicester, United Kingdom
              \email{mab@leicester.ac.uk} \and Astronomisches Rechen--Institut, Zentrum f\"{u}r Astronomie der
       Universit\"{a}t Heidelberg, 
       M\"{o}nchhofstr.~12--14, 69120 Heidelberg, Germany \email{jordan@ari.uni-heidelberg.de} }

   \date{Received ??, 2023; accepted ??, 2023}

 
  \abstract
   {The latest Gaia data release in July 2022, DR3, in addition to the refinement of the astrometric and photometric parameters, added a number of important data products to those available in earlier releases, including radial velocity data, information on stellar multiplicity and XP spectra of a selected sample of stars. Gaia has proved to be an important search tool for white dwarf stars, readily identifiable from their absolute $G$ magnitudes as low luminosity objects in the H-R diagram. Each data release has yielded large catalogues of white dwarfs, containing several hundred thousand objects, far in excess of the numbers known from all previous surveys ($\sim$ 40,000). While the normal Gaia photometry ($G$, $\Gbp$ and $\Grp$ bands) and astrometry can be used to identify white dwarfs with high confidence, it is much more difficult to parameterise the stars and determine the white dwarf spectral type from this information alone. Observing all stars in these catalogues with follow-up spectroscopy and photometry is also a huge logistical challenge with current facilities. }
   {The availability of the XP spectra and synthetic photometry presents an opportunity for more detailed spectral classification and measurement of effective temperature and surface gravity of Gaia white dwarfs.}
   {A magnitude limit of $G < 17.6$ was applied to the routine production of XP spectra for Gaia sources, which would have excluded most white dwarfs. Therefore, we created a catalogue of 100,000 high-quality white dwarf identifications for which XP spectra were processed, with a magnitude limit of  $G < 20.5$. Synthetic photometry was computed for all these stars, from the XP spectra, in Johnson, SDSS and J-PAS, published as the Gaia  Synthetic Photometry Catalogue - White Dwarfs (GSPC-WD). We have now taken this catalogue and applied machine learning techniques to provide a classification of all the stars from the  XP spectra. We have then applied an automated spectral fitting programme, with chi-squared minimisation, to measure their physical parameters (effective temperature and $\logg$) from which we can estimate the white dwarf masses and radii.}
   {We present the results of this work, demonstrating the power of being able to classify and parameterise such a large sample of $\approx 100,000$ stars. We describe what we can learn about the white dwarf population from this data set. We also explore the uncertainties in the process and the limitations of the data set.}
   {}

\titlerunning{Classification and parameterisation of Gaia white dwarfs}
\authorrunning{Vincent et al.}
   \maketitle
%

\section{Introduction}

The publication of the data from the ESA Gaia space mission \citep{GaiaCollaboration2016,GaiaCollaboration2018,GaiaCollaboration2021,GaiaCollaboration2023b} has revolutionized our understanding of white dwarfs by providing precise astrometric and photometric measurements, allowing for detailed studies of their properties, distributions, and evolutionary pathways.

\cite{GaiaCollaboration2018b}
has revealed a previously unseen division in the white dwarf sequence on the colour-magnitude diagram, providing unprecedented detail. Specifically, the Q branch and the A-B bifurcation have been identified. The Q branch is now believed to be caused by energy released during the crystallization of the white dwarf core \citep{Tremblay2019}, while the origin of the A-B bifurcation remained unclear until recently. The A branch primarily consists of white dwarfs with hydrogen-rich atmospheres, while the presence of the B branch has long been unexplained. \cite{El-Badry2018} attributed the existence of the B branch to a flattening in the initial-to-final-mass relation (IFMR), resulting in a secondary peak in the white dwarf mass distribution at approximately 0.8\msun. Soon after, \cite{Kilic2018} also suggested the presence of this secondary peak but attributed it to the occurrence of stellar mergers, while \cite{Bergeron2019} have shown that adding an invisible trace of hydrogen to the B branch stars caused their mass to move closer to the fiducial 0.6\msun. Only recently have \cite{Camisassa2023} and \cite{Blouin2023} been able to explain the B branch using white dwarf models that include helium and a small amount of carbon contamination, which cannot be detected optically and that the mass distribution of these white dwarfs is consistent with the mass distribution observed in hydrogen-rich white dwarfs and their standard evolutionary pathways. Furthermore, \cite{Blouin2023} have shown that neither the convective mixing of residual hydrogen nor the accretion of hydrogen or metals can be the dominant drivers of the bifurcation.

Gaia has uncovered a significant number of new white dwarfs, resulting in the identification of 73,000 white dwarfs within the 100-parsec solar neighbourhood in Gaia Data Release 2 \citep{GaiaCollaboration2021} by \cite{Jimenez-Esteban2019}. A total of 260,000 high-probability white dwarfs were discovered by \cite{GentileFusillo2019}, not restricting the search to the close Solar neighbourhood. The utilization of Gaia Early Data Release 3 \citep{GaiaCollaboration2021} by \cite{GentileFusillo2021} further increased the number of high-confidence white dwarf candidates to 359,000. This represents more than an order of magnitude increase compared to the previous number of known white dwarfs. As a result, several comprehensive analyses of Gaia white dwarfs belonging to various spectral classes were conducted, \citep[e.g.][]{Tremblay2019,Bergeron2019,Coutu2019,caron2023}.

The photometry provided by Gaia, from Data Release 1 \citep{GaiaCollaboration2016} to Early Data Release 3 \citep{GaiaCollaboration2021}, utilized wide passbands ($G$, $G_{\rm BP}$, and $G_{\rm RP}$), which are not ideal for precise model atmosphere modelling. Consequently, additional surveys employing narrower bands have been incorporated alongside Gaia data to enable more comprehensive analyses.

In Gaia Data Release 3 \citep{GaiaCollaboration2023b}, a significant advance was made by publishing more than 200 million low-resolution spectra from the Gaia blue (BP) and red (RP) photometers, covering a wavelength range of 330 nm $\leq \lambda \leq$ 1050 nm. These spectra, referred to as XP spectra, have provided invaluable data \citep{GaiaCollaboration2023a}. Among these XP spectra, nearly one hundred thousand are attributed to white dwarfs, offering a substantial resource for further analysis. 

The spectra obtained from Gaia Data Release 3 can be converted to synthetic photometry in physical units for any passband within the wavelength range they cover. This allows for direct comparisons with photometric data obtained by other surveys, such as the SDSS, and facilitates the interpretation of synthetic photometry derived from Gaia through model calculations specific to these passbands. Tables with synthetic photometry were published by Gaia Collaboration \citep{GaiaCollaboration2023a}, which includes standardized photometry for over 200 million sources across multiple widely-used photometric systems.

In order to leverage the full potential of the large data sets provided by the latest surveys, statistical algorithms have become essential in the field of astronomy. Machine learning tools for the analysis of white dwarfs have recently begun emerging, such as the \texttt{WDTools} package for spectroscopic parameter inference of DAs \citep{Chandra2020} and the classification pipeline of \cite{Vincent2023}, which contains several neural network classifiers for WD primary spectral type classification, as well as WD candidate and WD$+$MS binary system identification. The identification of WD$+$MS systems within Gaia has been studied in greater detail by \cite{Echeverry2022}, who tested the Random Forest classifier algorithm using realistic population of white dwarfs with XP-quality synthetic spectra. 

In this paper, we spectroscopically classify the white dwarfs of the GSPC sample using machine learning techniques and measure their stellar parameters. We describe the GSPC data in Section \ref{sec:DR3} and our classification methodology and results in Section \ref{sec:classifier}. We outline the parameterisation procedure in Section \ref{sec:param} and verify the consistency between the measured physical parameters and spectral classifications. We discuss selected results in Section \ref{sec:disc} and give our concluding remarks in Section \ref{sec:conc}.

\section{Gaia DR3 and the catalogue of synthetic photometry}\label{sec:DR3}

One of the new outputs from the Gaia DR3 catalogue, compared to DR2 and EDR3, was the publication of flux-calibrated low-resolution spectrophotometry for $\approx$ 220 million sources \citep{Gaia_DR3_summary}.
Gaia produces low resolution (R$\approx$50) spectra in each of its blue and red photometric channels, labelled BP and RP respectively.
Dispersion is provided by a prism. The raw spectra are calibrated and merged into a single XP spectrum, as described by \citet{DeAngeli2023} and \citet{Montegriffo2023}. The spectra are defined  as an array of coefficients to be applied to a set of basis functions.
Sampled spectra and synthetic photometry can be recovered from these coefficients using the GaiaXPy Python library (https://gaiaxpy.readthedocs.io/en/latest/description.html). An extensive survey of the possible uses of synthetic photometry from the Gaia XP spectra is presented in a paper accompanying the Gaia DR3 data release \citep{GaiaCollaboration2023a}.

In DR3 only mean spectra are available, the average from all valid observations included in the data processing pipeline. Furthermore, these are selected to have a reasonable number of observations (more than 15 transits) and to be sufficiently bright to ensure a good signal-to-noise ratio (S/N), generally limiting the $G$ magnitude to $ G\ <\ 17.65 $. However, if applied across the whole catalogue, this magnitude limit would exclude large numbers of interesting classes of fainter objects. Therefore, a few samples of specific objects that could be as faint as $ G\approx 21.43$ were added: including about 500 sources used for the calibration of the BP/RP data, a catalogue of about 100,000 white dwarf candidates, 17,000 galaxies, about 100,000 QSOs, about 19,000 ultra-cool dwarfs, 900 objects that were considered to be representative for each of the 900 neurons of the self-organising maps (SOMs) used by the outlier analysis (OA) module (see Sect. 9) and 19 solar analogues \citep{DeAngeli2023}.

The input catalogue of white dwarfs used to select suitable sources for generating XP spectra is described in detail in \cite{GaiaCollaboration2023a}. It derives from earlier work carried out to identify white dwarfs in the Gaia DR2 and eDR3 catalogues carried out by \cite{GaiaCollaboration2018b}, \cite{GentileFusillo2019, GentileFusillo2021} and the Gaia Catalogue of Nearby Stars \citep{Smart2021}. The input sample was drawn from the eDR3 release, designed to span the complete range and colours of white dwarfs, defined as high-probability white dwarf candidates by their location in the H-R diagram. The following criteria were applied:

\begin{itemize}
    \item{Equations 1-9 detailed in Gentile Fusillo et al. (2019)}
    \item{astrometric\_excess\_noise $< 5 $}
    \item{phot\_bp\_mean\_flux\_over\_error $<$ 20}
    \item{phot\_rp\_mean\_flux\_over\_error $<$ 20}
    \item{parallax/parallax\_error $<$ 10}
    \item{phot\_g\_mean\_flux\_over\_error $<$ 20}
    \item{${\rm log(parallax\_over\_error) < }$ \\ ${\rm 
    1.56(log(10^{3}/parallax)-3.17)+0.96}$}
\end{itemize}

\begin{figure}[ht]
    \centering
    \includegraphics{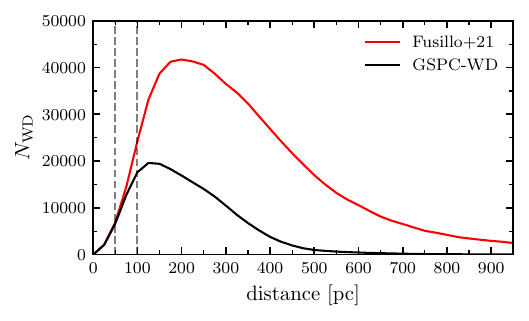}
    \caption{Number of objects in the \cite{GentileFusillo2021} white dwarf candidate catalogue with $P_{\rm WD}>0.65$ (red line) and number of objects in the GSPC-WD catalogue (black line) as a function of distance. The grey dashed lines indicate distances of 50 and 100 pc.}
    \label{fig:completeness}
\end{figure}

This yields a sample of 102,000 white dwarfs, five times the number of known spectroscopically identified objects. Most of the objects have $G < 19.5$, but about 30\% \ are fainter. The effective $G$ magnitude cut-off is $\approx 20$. This catalogue corresponds to the list of objects included in the Gaia Synthetic Photometry Catalogue for White Dwarfs (GSPC-WD). The completeness of the catalogue essentially follows the same trends as those in \cite{GentileFusillo2021} up to $\sim$50 pc, after which the number of XP objects starts diminishing. To illustrate this, we plot in Figure \ref{fig:completeness} the number of white dwarf candidates with $P_{\rm WD}>0.65$ from the \cite{GentileFusillo2021} along with the number of objects in the GSPC-WD catalogue as a function of distance. For reference, the 50 pc and 100 pc distances are highlighted with grey dashed lines. We also looked for a color-dependent drop in completeness noticed by \cite{Jimenez-Esteban2023}, who analyzed the 100 pc sample of the GSPC-WD catalogue and found a gradual decrease of completeness with redder $\Gbp-\Grp$ colors. We plotted the ratio between the number of objects in the GSPC-WD catalogue and objects in the \cite{GentileFusillo2021} white dwarf candidate catalogue with $P_\mathrm{WD}>0.75$ against $\Gbp-\Grp$ color bins (not shown here), and found a small decrease of GSPC-WD objects around $\Gbp-\Grp=0.2$ that remained nearly constant for redder colors. 

\begin{figure*}[ht]
    \centering
    \includegraphics{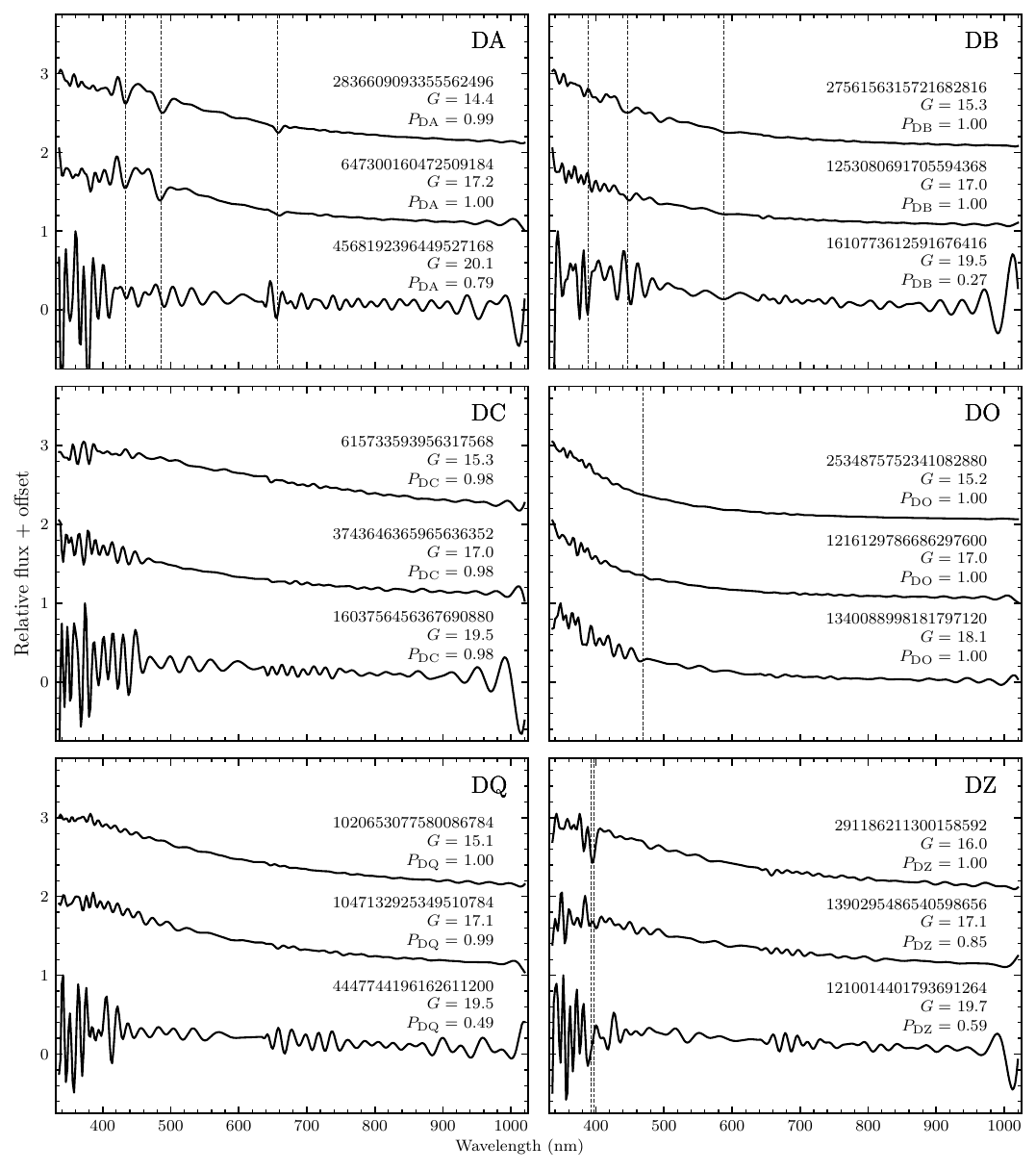}
    \caption{Gallery of XP spectra for Gaia objects with SDSS-confirmed spectroscopic type (indicated in the top right corner of each panel). Three spectra of varying $G$ magnitudes are displayed to illustrate the difference between brighter and fainter stars. Spectral lines typically used for classification are displayed for reference (see text). The Gaia DR3 source identification and the spectroscopic type probability predicted by our classifiers ($P_\mathrm{class}$) are also shown.}
    \label{fig:XPgallery}
\end{figure*}

Figure \ref{fig:XPgallery} shows examples of typical XP spectra, generated from the coefficients, for the six primary white dwarf spectral types: DA, DB, DC, DO, DQ and DZ, spanning a range of magnitudes. These objects are previously known white dwarfs with spectral types confirmed by SDSS. The strongest H Balmer lines ($\lambda6562 , \lambda4861 , \lambda4340$) are visible in the DA spectra, while He I lines ($\lambda5876 , \lambda4471 , \lambda3889$) and and the calcium lines ($\lambda3934 , \lambda3969$) are identifiable in the DBs and DZs, respectively. The He II $\lambda4686$ line typically used for classification of DOs appears, however, invisible at this resolution. 
The classifiers can, nevertheless, make confident predictions by using alternative information such as the stellar continuum shape, as the XP coefficients are not processed in such a way that it is removed. The lowest S/N regions correspond to the extremes of the wavelength range covered by the spectra, where the XP coefficients generate a pattern of oscillations due to the noise.

The following sections describe how we have applied machine learning to classify the GSPC-WD stars and determine their physical parameters using the synthetic colours derived from XP spectra. We then present the results we have obtained by studying this large sample of objects, a more than factor 5 increase in the number of white dwarfs of known spectral type.

\section{A machine learning approach to classification of the white dwarf XP spectra}\label{sec:classifier}
\subsection{Methodology}
Spectral classification of white dwarf XP spectra in the Gaia XP sample is conducted using a "one-versus-all" approach. This method employs binary classifiers that are trained to distinguish a specific class from all other white dwarfs. To perform the classification, we utilize the Scikit-learn \texttt{GradientBoostingClassifier} \citep{scikitlearn}, which is an ensemble method that trains multiple regression trees by minimizing a differentiable loss function and then combines them into a powerful model. Further details regarding the hyperparameters employed and data preprocessing can be found in Appendix \ref{appendix:classifier}.

Our classifiers take the 110 XP coefficients as input and provide a probability ranging from 0 to 1, indicating the likelihood of a spectrum belonging to a particular class. The spectral type with the highest probability is assigned to the object if it surpasses a minimum confidence threshold, as discussed in the subsequent section. In our study, the classifiers are trained to distinguish six primary spectral types: DA, DB, DC, DO, DQ, and DZ. It is important to note that our classification scheme solely accounts for the primary spectral type and does not attempt to identify secondary signatures. For instance, in our approach, a white dwarf spectrum exhibiting both hydrogen and neutral helium lines would be classified as DA or DB based on the relative strengths of these lines. Conversely, in the traditional system proposed by \cite{sion1983}, the spectrum would be classified as either DAB or DBA, where the spectral type order also depends on the relative strengths of the lines.

The training data utilized in our study were obtained from the Gaia-SDSS catalogue described in \cite{Vincent2023}. This catalogue provides a robust data-driven classification of primary spectral types for 27,866 unique Gaia white dwarfs, employing spectra from the SDSS Data Release 17 \citep{abdurrouf2022}. The catalogue assigns a probability $P_\mathrm{class}$ to indicate the likelihood of a spectrum belonging to one of the following 13 classes: DA, DB, DC, DO, DQ, hotDQ, DZ, DAH, PG1159, cataclysmic variable, sdB, sdO, or sdBO. As recommended in \citet{Vincent2023}, we restricted our selection to objects with a classification confidence $P_\mathrm{class}>0.6$ to ensure reliable spectroscopic classifications for our training data. Additionally, we required objects to possess at least one SDSS spectrum with a signal-to-noise ratio above 9 to further enhance the reliability of the spectroscopic classifications. These criteria resulted in a dataset consisting of 13,743 unique Gaia objects with spectroscopic classification. Table \ref{tab:classifier} displays the number of objects for each of the six primary spectral types used to train and validate the XP classifiers.

To validate our choice of classification algorithm, we employed cross-validation. For each class, we trained 20 gradient boosting classifier models using distinct data splits. Each split comprised a random selection of 10\% of the data for the test set, 10\% for the validation set, and 80\% for training. We assessed the mean precision and recall scores at a classification confidence threshold of 0.6 on the test set of the 20 models, which are presented in Table \ref{tab:classifier} as $P_\mathrm{c-v}$ and $R_\mathrm{c-v}$, respectively. The overall performance is acceptable, with every class achieving precision and recall scores above $\gta60$\%. Although more complex classifier models might yield slightly higher scores, the limited number of objects (less than 1000) for most classes increases the risk of overfitting. Furthermore, there is no guarantee that the spectral features used to classify the objects when employing SDSS spectra remain discernible in the lower-resolution XP spectra. This situation could lead to complex models focusing excessively on learning features that do not exist. Consequently, a certain amount of error is expected in the training labels due to the invisibility of spectral features at low resolution for certain objects.

Examining individual spectral types, the DA and DB classifiers exhibit excellent performance, with both precision and recall values exceeding 95\%. This outcome can be attributed to the larger number of objects in these classes and their easily distinguishable spectral features. The DC classifiers display the poorest performance, although still acceptable, with approximately 66\% precision and recall. Spectral features used to classify SDSS spectra may become invisible, causing many objects labeled as non-DC to be classified as DC at XP resolution. Conversely, objects labeled as DC are highly likely to possess correct labels, as no new features become visible when the resolution is downgraded. This assurance allows us to consider high-confidence DC classifications as genuine. The DQ and DZ classifiers demonstrated good recall performance of around 85\%, but relatively low precision of approximately 60\%. The lower performance for these spectral types is anticipated, as the spectral features typically used to classify them, such as Swan bands or other carbon absorption lines for DQ white dwarfs and $\mathrm{Ca}~\mathrm{II}$ absorption features for DZ \citep{kleinman2013, Coutu2019, kepler2021}, become increasingly challenging to distinguish in low-resolution and/or low signal-to-noise spectra. This difficulty is evident in the XP spectra gallery displayed in Figure \ref{fig:XPgallery}. DO white dwarfs also encounter this issue, with the $\mathrm{He}~\mathrm{II}$ $\lambda4686$ line being virtually invisible in most spectra. The DO cross-validation scores for precision and recall were both 76\%, which may seem acceptable at first glance. However, these scores should be interpreted with caution due to the limited number of objects. The DO classifiers are extremely sensitive to the choice of training and testing data, resulting in individual model performance scores fluctuating between 60\% and 100\%.

To mitigate the biases learned by individual models, we performed ensemble learning by combining the top-5 performing models of each class based on their test F-scores, i.e. the average of precision and recall on the test set. We obtained the final class probability by averaging the predictions of the ensemble models. Table \ref{tab:classifier} presents the average precision and recall scores of the top-5 ensembles at a threshold of 0.6 for comparison with the cross-validation scores. Ensembling the top-5 models yielded a minor improvement of a few percentage points over the cross-validation results, except for the DO classifiers, which improved by approximately 20\%. This improvement highlights the performance variance resulting from the limited number of objects. The top-5 ensembles were employed to classify the GSPC-WD sample, as discussed in the subsequent section.

\input{tables/classifier2}

\begin{figure}[ht]
    \centering
    \includegraphics[width=1.\columnwidth]{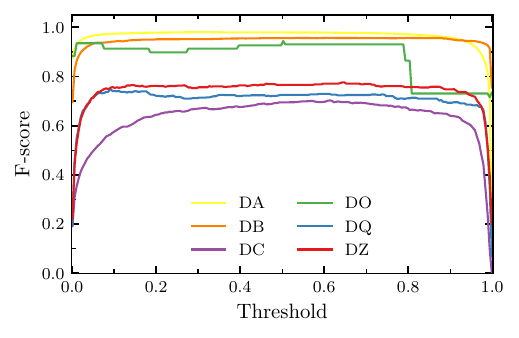}
    \caption{F-score curves of the top-5 performing classifier ensembles. A threshold value of 0.65 maximises the mean F-scores for all classes.}
    \label{fig:Fbeta}
\end{figure}

\subsection{Classification of the GSPC-WD sample}
We started by refining the selection of our initial 102,000 WD candidates following \cite{Vincent2023}, which employed neural networks and 13 Gaia parameters to estimate the probability of an object being a white dwarf ($P_{WD}$) for approximately 1.3 million Gaia objects described in Fusillo et al. (2021). Their machine learning approach produced superior results compared to density estimation on the Gaia Hertzsprung-Russell (H-R) diagram, particularly in regions where the white dwarf locus and main sequence stars overlap. We select high-confidence white dwarf candidates with XP spectra by applying a $P_{WD}>0.9$ cut and a standard deviation limit of 0.02 on $P_{WD}$, retaining a total of 100,886 objects.

To spectroscopically classify the GSPC-WD sample, we employed the top-5 classifier models for each spectral type, as described in the previous section. The mean of their predictions was used as the final classification probability. We determined the optimal confidence threshold for each ensemble by plotting the mean F-scores against threshold values as shown in Figure \ref{fig:Fbeta}. A threshold of approximately 0.65 maximizes the F-score for all classes. Objects with probabilities below 0.65 for any spectral type were still classified according to the most probable class but labeled as uncertain by adding a colon annotation (e.g., "DA:"). Feeding the 101,783 objects to the classifier ensembles, we obtain 89,188 high-confidence classifications, while 11,698 objects remained with uncertain classifications. The number of high-confidence classified objects per spectral type is presented in Table \ref{tab:GSPCWD}. The uncertain objects primarily constitute the fainter end of our sample, with an average $G$ magnitude of 18.9. Lowering the confidence threshold to 0.5 provided classifications for an additional 4745 objects, but caution should be exercised as this may introduce a larger number of false positives. A more pure but incomplete sample may be selected by increasing the confidence threshold or, vice versa, a more complete but less pure sample by decreasing the threshold. The probability $P_\mathrm{class}$ for an object belonging to any of the six possible classes is available in our online catalog described in Section \ref{sec:catalogue}.

From Table \ref{tab:GSPCWD}, we see that the relative number of objects for each class is similar to what is found within the Gaia-SDSS catalogue of \cite{Vincent2023}, which provides some assurance as to the global performance of the classifiers, but is not unexpected. As pointed out in \cite{BailerJones2019}, most classification algorithms implicitly learn a prior probability for each class based on their relative proportions in the training data. We verified this by calculating the ratio of each class (number of objects per class, including low-confidence classifications, over all XP objects) and found nearly exactly the same ratios as those obtained for each class in the training dataset (see Table \ref{tab:classifier}). Although the SDSS currently provides the largest spectroscopic sample of white dwarfs, and thus the best available prior, the completeness of the XP-SDSS white dwarf sample as well as the SDSS target selection biases should be taken into consideration when doing population analyses requiring high degrees of statistical precision.

\begin{figure*}[ht]
    \centering
    \includegraphics{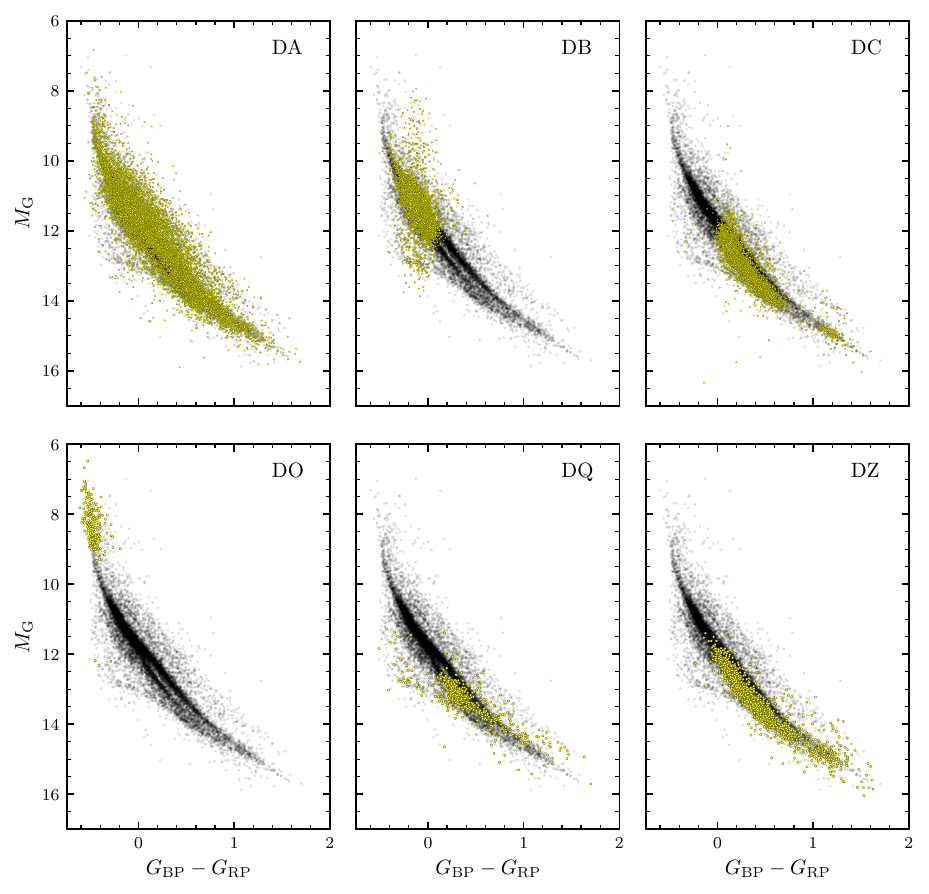} 
    \caption{H-R diagram of GSPC-WD objects with high classification confidence and converged fits. The location of objects for every class is consistent with previous spectroscopic studies. For clarity, background objects (grey points) are restricted to those with $G<18$ and a parallax measurement error less than 1\%, and only random selection of 25\% of all DA is shown.}
    \label{fig:HRD}
\end{figure*}

We perform a brief sanity check to verify whether the classifications follow established trends when visualized with data other than the XP spectra themselves. To this end, we plot the high-confidence classification results on the Gaia H-R diagram in Figure \ref{fig:HRD}. The locations of classified objects are consistent with expectations: DO stars are found at the hot end of the white dwarf locus \citep{bedard2020}, DA stars mainly populate the A-branch \citep{Bergeron2019}, DB stars are located on or near the warmer section of the B-branch, while DQ and DZ stars are situated on the cooler section of the B-branch \citep{Coutu2019} along with DC stars, the latter which are also found at the faint end of the white dwarf locus. An alternative view of the classifications is shown on the bottom panel of Figure \ref{fig:grug}, where we plot the synthetic SDSS $(g-r)$ vs $(u-g)$ colour-colour diagram for objects with both XP spectra and SDSS photometry, overlaid with the theoretical colour tracks for pure hydrogen and pure helium at a constant mass of 0.6 $M_\odot$. While the XP spectra and their synthetic photometry are not independent, the DA and non-DA classifications fall nicely on the appropriate tracks, providing further assurance that the appropriate atmosphere models are used to measure their stellar parameters.

Overall, the reliability of our spectroscopic classification for the 100,886 GSPC-WD objects is supported by reasonable relative class numbers, as well as their location on the H-R and color-color diagrams. In the following section, we describe the atmosphere models and procedures used to measure the stellar parameters of this large sample.

\input{tables/GSPC-WD}

\section{Parameterisation of the GSPC-WD sample}\label{sec:param}
\subsection{The Photometric Technique}\label{sec:photo}
We measure the physical parameters in our sample using the so-called photometric technique described in \cite{Bergeron1997}. Briefly, the standardized synthetic SDSS magnitudes are converted into average fluxes using appropriate zero-points \citep{Montegriffo2023} and conversion equations \citep{Holberg2006}. Model photometry is then calculated for class-specific model grids by integrating the monochromatic Eddington fluxes over each bandpass. These model fluxes depend on the effective temperature $T_\mathrm{eff}$, the surface gravity $\logg$ and chemical composition. The observed and model fluxes are then related to each other via the solid angle $\pi(R/D)^2$, where $R$ is the stellar radius and $D$ is the distance from Earth. Since the distance is known from Gaia parallax measurements, the radius can be measured directly and converted into stellar mass $M$ using evolutionary models, which provide a temperature-dependent mass-radius relation. We rely on the evolutionary models described
in \cite{bedard2020} with C/O cores, $q({\rm He})\equiv \log M_{\rm
  He}/M_{\star}=10^{-2}$ and $q({\rm H})=10^{-4}$, which are
representative of H-atmosphere white dwarfs, and $q({\rm He})=10^{-2}$
and $q({\rm H})=10^{-10}$, which are representative of He-atmosphere
white dwarfs\footnote{The models can be found here: \url{https://www.astro.umontreal.ca/~bergeron/CoolingModels}}. A chi-squared minimization is performed between the observed and model average fluxes using the method of Levenberg-Marquardt \citep{Press1986}. In our fitting procedure described below, $T_\mathrm{eff}$ and $\pi(R/D)^2$ are set as free parameters while the chemical composition remains fixed.

The synthetic SDSS magnitudes are dereddened using the 3D extinction maps and parameterisation in \cite{GentileFusillo2021} and we apply the parallax zero-point correction described in \cite{Lindergren2021}. Moreover, it is well established that the SDSS magnitude system is not exactly on the AB magnitude system \citep[see][ and references therein]{Bergeron2019}, requiring $uiz$ magnitude corrections proposed by \cite{Eisenstein2006}, which we apply. Since the given errors on the synthetic SDSS fluxes tend to be smaller than the AB system corrections, we follow \cite{Bergeron2019} and adopt a lower limit of 0.03 mag uncertainty in all bandpasses.

\begin{figure}[ht]
    \centering
    \includegraphics[width=0.95\columnwidth]{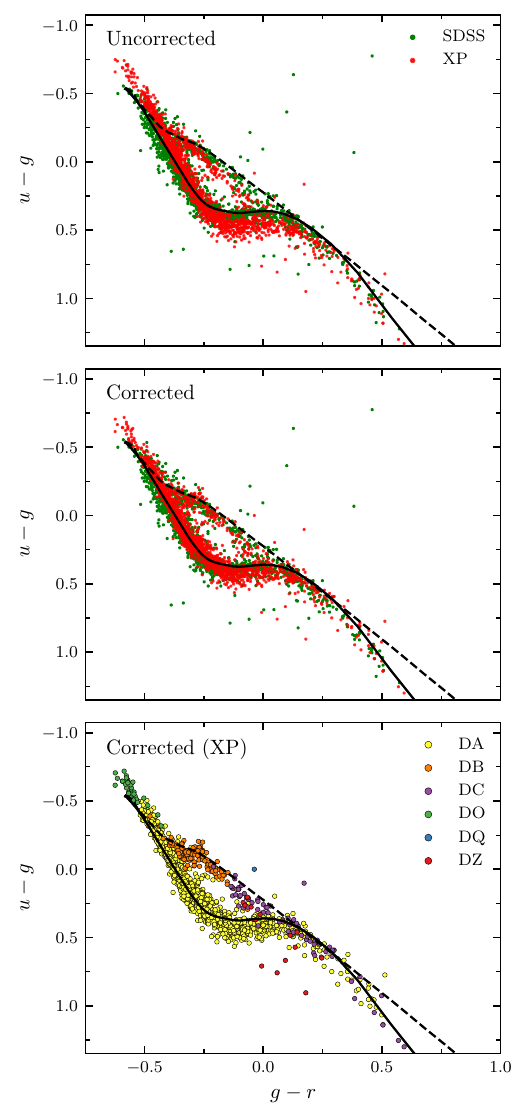}
    \caption{Color-color diagrams of objects with both real and synthetic SDSS photometry. For clarity, only objects with $G<17$ are shown. The top panel contains the two color-color distributions before the $u$ band correction (see text), displaying a significant color-dependent shift between the two. The middle panel shows the distributions after the $u$ band corrections have been applied to the synthetic photometry. The bottom panel shows the corrected synthetic photometry color-color diagram with points colored by their XP spectroscopic classification. On all three panels, the cooling tracks for pure hydrogen (full black line) and pure helium (dashed black line) are shown at a constant mass of 0.6\msun.}
    \label{fig:grug}
\end{figure}

\begin{figure}[ht]
    \centering
    \includegraphics{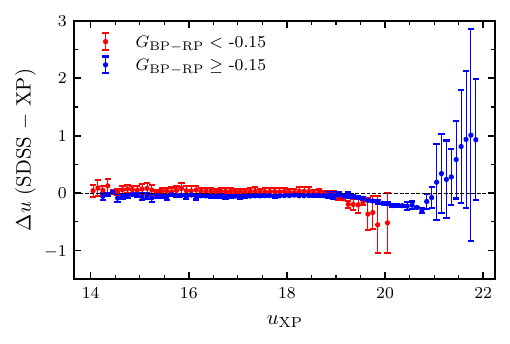}
    \caption{Median difference between real and synthetic SDSS photometry for bins of 0.2 synthetic $u$ magnitudes. Error bars correspond to the 67.5th percentile of each bins.}
    \label{fig:ucorr}
\end{figure}

Ultraviolet fluxes of XP spectra, and, therefore, synthetic ultraviolet bandpasses, are known to have strong color-dependent systematic errors. We refer the readers to \cite{Montegriffo2023} and \cite{Montegriffo2023} for a lengthy analysis of these issues. While the standardized synthetic photometry (see Section \ref{sec:catalogue}) tries to provisionally address this issue, it remains a blanket fix for the entire GSPC sample and is not perfectly adapted to white dwarfs. We find a significant color-dependent shift persists in the $u$ band and attempt to further reduce it. We crossmatch our GSPC-WD catalogue with SDSS DR18 photometry and find 21,254 objects in common. The topmost plot in Figure \ref{fig:grug} illustrates this shift in the $u-g$ vs. $g-r$ color-color diagrams between real and synthetic SDSS photometry. As can be seen from the plot, synthetic $u$ magnitudes tend to be overestimated for colors $g-r\ \lta -0.25$ and overestimated at $g-r\ \gta -0.25$. We found that this split roughly corresponds to the Gaia color $\Gbp-\Grp=-0.15$ and use this as a separation point to calculate a $u$ magnitude correction for blue ($\Gbp-\Grp<-0.15$) and red ($\Gbp-\Grp\geq -0.15$) white dwarfs. We calculate the correction term for each color group by binning white dwarfs into synthetic $u$ bins of 0.02 mag and computing the median difference between real and synthetic magnitudes. The median difference, along with the 67.5th percentile as error bars, are shown as a function of synthetic $u$ magnitude in Figure \ref{fig:ucorr}. We then subtract this correction term from the synthetic $u$ magnitudes and use the 67.5th percentile as the uncertainty on the magnitude, as it is typically larger than the error provided by the XP pipeline. The effects of this correction on the $u-g$ vs. $g-r$ diagram is shown on the middle plot of Figure \ref{fig:grug}, where both the real and synthetic colors are now in better agreement. The correction term for each object in our sample can be found under the \texttt{u\_corr} and \texttt{u\_corr\_675} columns of our catalogue. Furthermore, the correction bins shown in Figure \ref{fig:ucorr} are made available as Supplementary Material.

\subsection{Model Atmospheres}\label{sec:atmo}
In this section we outline the model atmospheres used to measure the stellar parameters of the GPSC-WD sample and briefly discuss how unusual parameters may inform us about issues with the data or erroneous classification. Objects that have an uncertain classification were also fitted using the approaches described here based on their most probable class, although they are not discussed.

\subsubsection{DA White Dwarfs}\label{sec:DA}
Our model atmospheres for DA white dwarfs are described at length in \cite{Blouin2018a} for $\Te \leq 5000$ K, \cite{Tremblay2011} for $5000\ {\rm K} < \Te < 35,000$ K, and \cite{bedard2020} for $\Te \geq 35,000$ K. We assume a pure hydrogen atmospheric composition with model atmosphere grid parameters ranging from $3000\ {\rm K}\leq \Te \leq 150,000$ K and $6.5 \leq \logg \leq 9.5$ for the 77,330 DA stars, a sound assumption for the vast majority of DA white dwarfs \citep{Bergeron1997, Blouin2019}. The fitting procedure did not converge for 1080 objects, nearly all located in the hot tail of the WD locus or the WD+MS region on the Gaia H-R diagram. 248 of these objects have available spectroscopic classification and stellar parameters on the Montreal White Dwarf Database \citep[MWDD;][]{Dufour2017}, including 70 DA+MS binary systems, 1 CV, 1 DC, and 1 DAZ. The remaining 175 are all confirmed DA, out of which 127 are hot (\gta 30,000 K).

The stellar parameters of the hot DA stars in our sample require a note of caution. We find 34 DA with extremely high temperatures ($\Te> 300,000$ K), 14 of which are confirmed DAO or hot DA according to the MWDD. These temperatures are obviously implausible and are likely due to a combination of the $u$ band calibration issues discussed in Section \ref{sec:photo} and the sensitivity of optical photometry to highly effective temperatures. At $\Te> 40,000$ K, the spectral energy distribution of optical photometry is in the Rayleigh-Jeans regime and becomes a poor indicator of temperature. Figure 3 of \citet{bedard2020} illustrates this well, showing that the difference in $u-g$ colour between $\Te=50,000$ K and $\Te=100,000$ K for a 0.6\msun\; white dwarf is a mere 0.06 mag. One can also see from Figure \ref{fig:grug} that the offset between real and synthetic SDSS $u$ photometry is larger than 0.06 mag for objects located at the hot tail of the color-color diagram, even when corrected, and keeps increasing for hotter temperatures. Another consequence of these issues is the trend of increasing mass with temperature, starting around 40,000 K, on the DA mass-$\Te$ diagram displayed in Figure \ref{fig:Mteff}. A slight error in the photometry can cause overestimation of the temperature, which in turn is compensated by underestimation of the stellar radius, leading to overestimation of the mass and the observed diagonal pattern. Precise measurement of the stellar parameters of hot white dwarfs would require UV or spectroscopic observations and are beyond the scope of this paper.

\begin{figure*}[ht]
    \centering
    \includegraphics{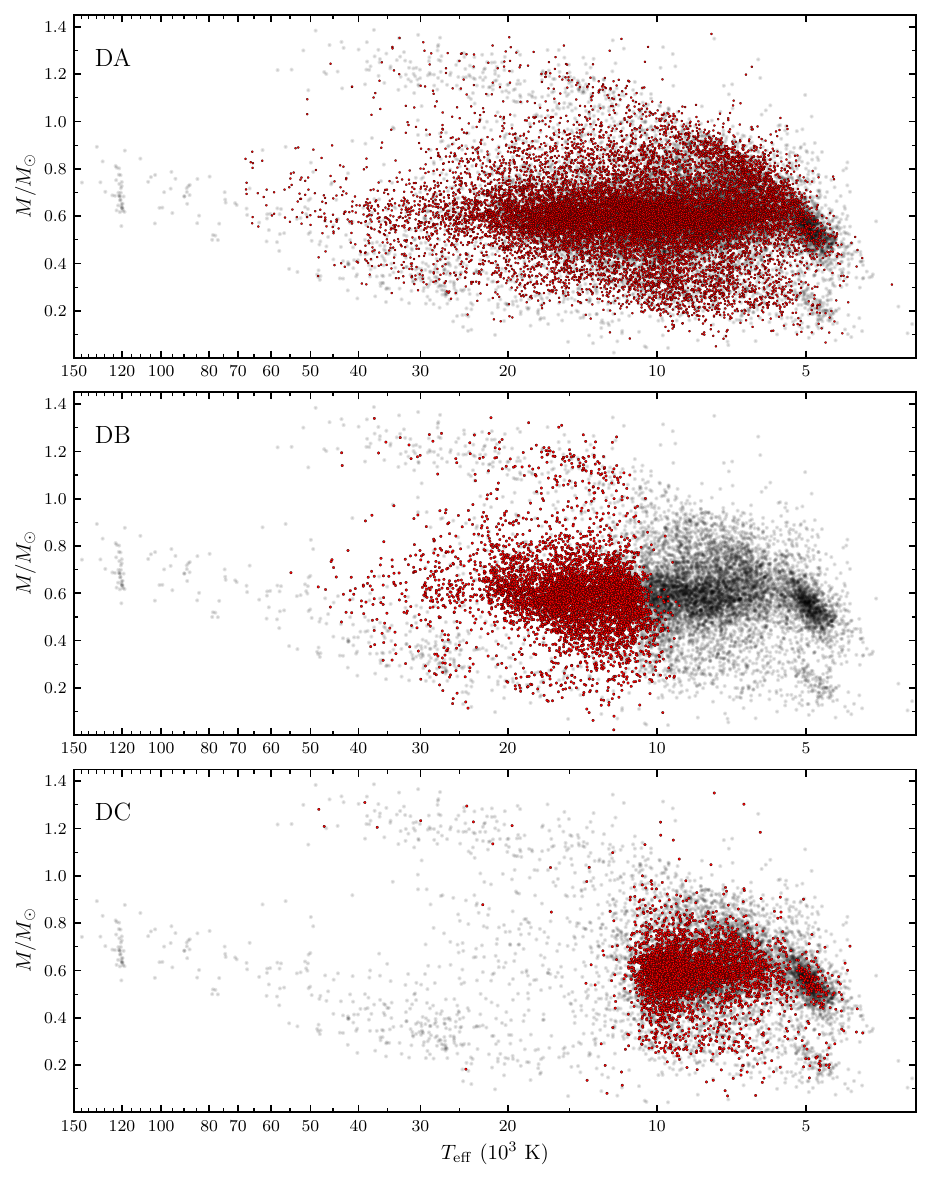}
    \caption{Mass-effective temperature diagrams for the six main spectroscopic types in the GSPC-WD sample. For clarity, the background objects (gray dots) are a fixed random selection of 50\% of all GSPC-WD objects, from which objects from the class being presented are excluded, and only 20\% of all DA are shown.}
    \label{fig:Mteff}
\end{figure*}

\begin{figure*}[ht]
    \addtocounter{figure}{-1} 
    \centering
    \includegraphics{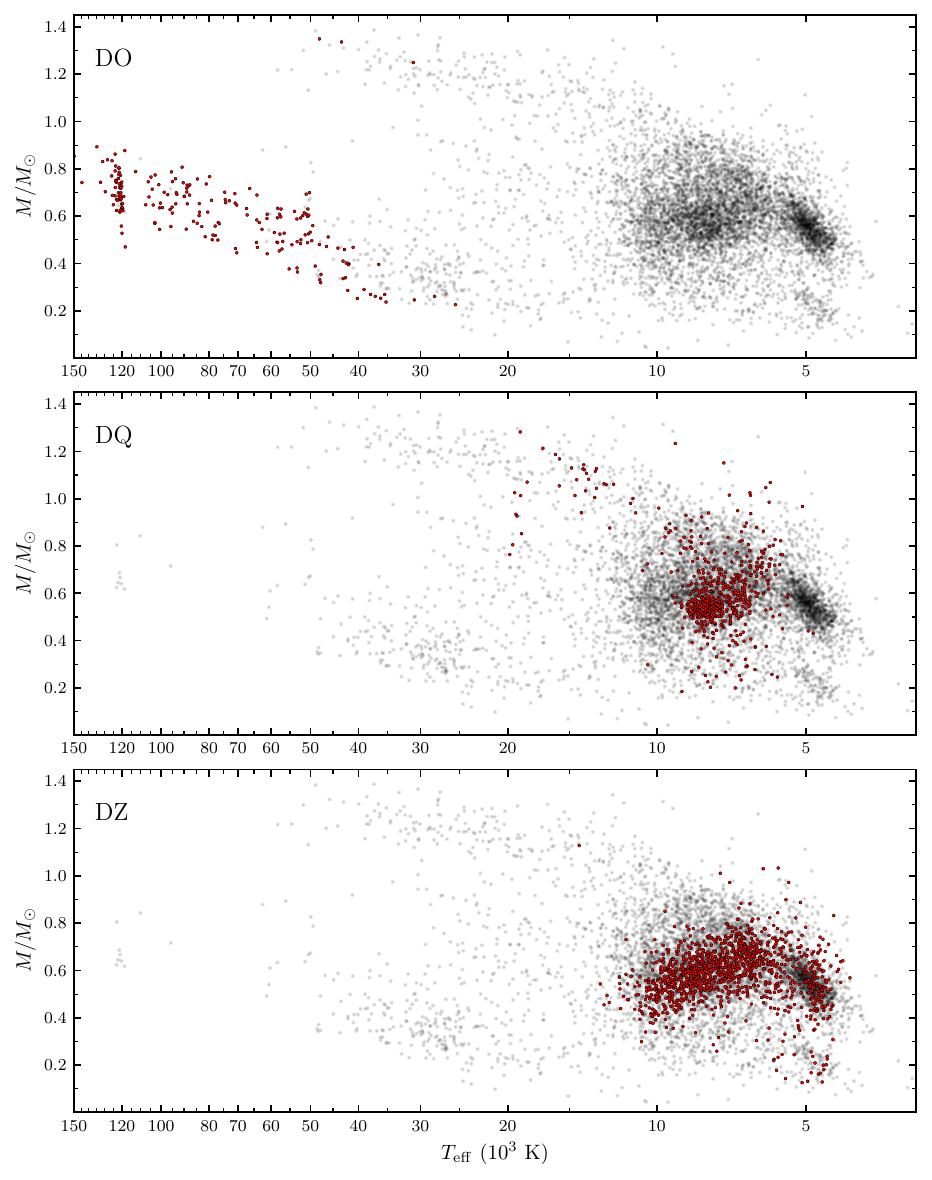}
    \caption[]{{\bf Continued.}}
\end{figure*}

We also note 60 DA with masses above 1.44\msun. These include the 34 objects with extremely high temperatures mentioned above, one confirmed cool DC analyzed in \cite{caron2023} and 30 objects scattered around the H-R diagram with no other spectroscopic observations available. We suspect the latter are either erroneous classifications or mixed-type white dwarfs with visible hydrogen in their atmosphere.

\subsubsection{DB White Dwarfs}\label{sec:DB}
For the 5688 DB white dwarfs, we use the model atmospheres described in \cite{Blouin2019} for $\Te\leq8000$ K, \cite{GBB2019} for $8000\ {\rm K}<\Te<40,000$ K, and \cite{bedard2020} for $\Te \geq 40,000$ K. We assume a pure helium atmosphere with grids covering $5000\ {\rm K}\leq \Te \leq 60,000$ K and $7.0 \leq \logg \leq 9.0$. The fitting procedure did not converge for 14 objects, one located in the hot tail of the WD locus and the rest in the WD$+$MS region of the Gaia H-R diagram. The MWDD has spectroscopic types available for two objects, indicating PG1159 for the hot object and DA+MS for the second one.

A surprising outcome of these results are 165 DB with masses above 1\msun, strongly at odds with previous in-depth analysis of large DB samples suggesting that massive DB are essentially nonexistent \citep{GBB2019, Bergeron2011}. Our massive DB are all located under the main white dwarf sequence, where the Q-branch and magnetic white dwarfs are found. Indeed, 13 of these massive DB have spectroscopic types available in the MWDD, 6 of which are classified as magnetic white dwarfs, 5 as DQ, 1 as DBA and 1 as DB:$+$MS. The high masses are likely to be genuine since high masses are common among magnetic \citep{Hardy2023a, Hardy2023b} and DQ white dwarfs found within the Q-branch \citep{Cheng2019, Coutu2019}. Visual inspection of the sampled XP spectra reveals that most of these objects have spectral features at the position of the He I lines at the bluest part of the spectrum, but also near the ionized carbon line $\lambda4267$. At such low resolution and signal-to-noise ratio, however, it is impossible to determine whether these features are truly single lines or magnetically distorted lines.

\subsubsection{DC White Dwarfs}
The determination of the atmospheric composition of DC white dwarfs is difficult because, by definition, they do not have any discernible features in their spectra. Furthermore, invisible traces of hydrogen and other heavy elements in helium-atmosphere DC can significantly affect the effective temperature and mass measurements when using the photometric technique \citep{Bergeron2019, Blouin2019, Blouin2023}. The exact amount of hydrogen is impossible to determine accurately, except for cool DC for which molecular hydrogen starts to form and causes strong infrared absorption via collision-induced absorption \citep{Bergeron2022, caron2023}. It is, however, possible to distinguish helium and hydrogen atmospheres due to the different behaviour of continuum opacities, but this requires precise photometric measurements and accurate magnitude-to-flux conversion. Considering the SDSS photometry is known to have calibration issues \citep{Holberg2006, Bergeron2019}, combined with the statistical error on the synthetic magnitudes \citep[a few mmag for $griz$ bands, see Section 3.1 of ][]{Montegriffo2023} and $u$ band calibration issues (see previous subsection), it is dangerous to let the H/He abundance ratio vary as a free parameter. We instead keep it fixed and follow the heuristic approach described below.

To measure the physical parameters of our 4082 DC stars, we use the same model atmospheres as those listed in Section \ref{sec:DA} and use different atmosphere compositions depending on the temperature based on the in-depth analysis of DC stars by \cite{caron2023}. We initially assume a pure helium atmosphere to have a rough estimate of the temperature, then fit the objects again according to this estimate. We assume a pure hydrogen atmosphere for stars under 5500 K, a mixed atmosphere with a fixed H/He=10$^{-5}$ abundance ratio for stars between 5500 and 11,000 K, and retain the pure helium atmosphere for the remaining hot objects.

The fitting procedure did not converge for 50 objects, which are mostly located at the end of the DC sequence on the B-branch and in the WD$+$MS region. 19 of these objects have spectroscopic classifications available in the MWDD, including 18 DC and one DQpec. We also note one DC, Gaia DR3 1505825635741455872, located in the so-called ultracool sequence \citep{Kilic2020, Bergeron2022} with a mass above 1.44\msun. This cool DC has been previously analysed by \cite{caron2023} who found a high, but more reasonable, mass of 1.18\msun.

\subsubsection{DO White Dwarfs}
To fit our 215 DO stars, we use the same model atmospheres described in \cite{bedard2020}. We assume a pure helium atmosphere, a suitable composition for most DO white dwarfs \citep{bedard2020}, and use model grids covering $30,000\ {\rm K}\leq \Te \leq 150,000$ K and $6.5 \leq \logg \leq 9.5$. The fitting procedure did not converge for 12 objects, 5 of which are confirmed DO(A) according to the MWDD.

Just as for hot DA, the precise measurement of physical parameters of DO stars is impossible. To briefly summarize the explanations in Section \ref{sec:DA}, optical photometry is extremely sensitive to effective temperature above 40,000 K and the $u$ band is known to have large systematic errors, causing the measured physical parameters to be unreliable for hot white dwarfs. As a matter of fact, the diagonal trend in the mass-$\Te$ caused by errors in the $u$ band is glaringly obvious for the DO mass-$\Te$ diagram in Figure \ref{fig:Mteff}. Additionally, a vertical pattern on the mass-$\Te$ diagram can be seen around 150,000 K, indicating that numerous objects have hit the upper temperature limit of the model grids during the fitting procedure. Given the situation, we advise against using the physical parameters we measured for DO stars as well as hot DAs for any analysis of these stars. These results should instead be interpreted as an illustration of the current limitations of the GSPC-WD sample.

Apart from the issues described above, we note three DOs with particularly dubious physical parameters. The three objects are located within the Q-branch, have very high masses ($>1$\msun) and low temperatures ($\sim$30,000 K to 50,000 K). They can be easily discerned from the H-R diagram in Figure \ref{fig:HRD} and DO mass-$\Te$ diagram in Figure \ref{fig:Mteff}. Only one of these objects has a spectroscopic classification in the MWDD and is a DC. Ionized helium absorption lines only appear above $\gta50,000$ K and are unlikely to be distinguishable at the resolution of XP spectra. Visual inspection of the spectra shows features near the positions of He {\sc ii} absorption lines, though identifying the spectral type remains difficult. As for the massive DB (see Section \ref{sec:DB}), we suspect magnetic white dwarfs whose distorted features happen to be confused with DO stars at low resolution.

\subsubsection{DQ White Dwarfs}
Recent spectroscopic analyses have clearly confirmed the existence of two distinct DQ evolutionary sequences: one with normal-mass white dwarfs and one with heavily carbon-polluted and generally more massive objects \citep{Dufour2005, Coutu2019, Blouin2019}. Since the DQ designation in our sample includes all carbon-contaminated atmosphere white dwarfs and does not differentiate between the two sequences, and since we do not have higher resolution spectroscopy to constrain the atmosphere carbon abundance, we employ a two-step fitting strategy along with empirical relations between temperature and carbon abundance in order to obtain the best possible stellar parameters.

The empirical relations between temperature and carbon abundance are made using the data from Figure 12 of \cite{Coutu2019}. We split the objects on their figure into two groups: one including high-mass and high-temperature objects ($M\geq0.7$\msun, $\Te \geq9000$ K) and one including the normal-mass and lower temperature objects ($M<0.7$\msun, $\Te <9000$ K). We then fit a linear curve to each group to predict the carbon abundance as a function of temperature.

Using the models of \cite{Coutu2019}, see also \cite{Blouin2018a, Blouin2019} and \cite{BlouinDufour2019}, we create two grids, one for warmer and one for cooler temperature ranges. More precisely, the warm grid covers $7 \leq \logg \leq 9$, $-5 \leq \log \mathrm{C/He} \leq -1$, $8000\ {\rm K} \leq \Te \leq 16,000$, and the cool grid covers $7.5 \leq \logg \leq 9$, $-7.5 \geq \log \mathrm{C/He} \geq -5$, $5000\ {\rm K} \leq \Te \leq 9000$ K. We assume there is no hydrogen in the atmosphere. We first fit all DQs assuming a fixed carbon abundance $\log \mathrm{C/He}=-5$ and leaving $\Te$ and $\logg$ as free parameters using both grids. We then select the best fit between the two grids based on the $\chi^2$ values and estimate a new carbon abundance using the appropriate empirical $\Te$-$\log \mathrm{C/He}$ relation. We then fit the photometry again using the best-fitting grid of the previous iteration along with the new carbon abundance as a fixed parameter.

Among our 601 DQ white dwarfs, the fitting procedure did not converge for 23 stars. Among the latter, 2 are located under the main WD locus of the Gaia H-R diagram while the rest are found in the WD$+$MS region. Spectroscopic types are available on the MWDD for 6 of them, including 4 CV, 1 hot DQ and 1 DAH. Some Hot DQ and DAH white dwarfs can be visually difficult to distinguish from one another, even for machine learning algorithms classifying SDSS spectra \citep{Vincent2023}. Extra steps should thus be taken in order to identify possible DAH misidentified as DQ at XP resolution for any detailed analysis of our DQ sample. As for the 4 CV, we visually inspect their spectra and find that one of them looks like a DC, while the other 3 show a strong absorption feature near the ionized carbon line $\lambda4267$. We inspect the SDSS spectra of these three objects and find no clear carbon absorption line at that wavelength. We suspect the "absorption feature" seen on the XP spectra is not an actual line, but appears as such at low resolution due to the contrast of being surrounded by two large emission lines.

\subsubsection{DZ White Dwarfs}
Since the training data of the DZ spectral type classifier is constructed in such a way to exclude most spectra with secondary signatures (e.g., DZA/DAZ and DZB/DBZ), we expect the majority of XP spectra classified as DZ to be white dwarfs with cool helium-rich atmospheres. This is supported by the fact that most objects classified as DZ lie on the B-branch of the Gaia H-R diagram as shown in Figure \ref{fig:HRD}. We use the model atmospheres described in \cite{Blouin2018a} and \cite{Coutu2019}, and assume a helium-dominated atmosphere with a fixed amount of metal pollution. Our solutions are provided in terms of the Ca abundance ($\log \mathrm{Ca/He}$), and we assume chondritic abundance ratios with respect to Ca for other metals. The model grids cover $7 \leq \logg \leq 9$ and $4000\ {\rm K} \leq \Te \leq 16,000$ K, while the calcium abundance ratio remains fixed at $\log \mathrm{Ca/He}=-9.5$. Although calcium abundance has been shown to vary with temperature \citep{Hollands2017, Coutu2019, BlouinXu2022}, the scatter of the Ca/He-$\Te$ relationship is very large and properly constraining the abundance requires higher resolution spectroscopy. A fixed assumption of $\log \mathrm{Ca/He}=-9.5$ is close to the mean abundance found by the previous studies and should be appropriate for the bulk of our DZ sample. 

We apply this procedure to our 1272 DZ stars and find 6 objects that do not converge, all located above the WD locus on the Gaia H-R diagram and none of which have spectroscopic analysis available in the MWDD.

As noted by \cite{Coutu2019}, the hydrogen-free atmosphere assumption results in slightly higher masses when applied to DZ stars (see their Figure 7). In order to shift the measured masses closer to the fiducial 0.6\msun, they included an invisible trace of hydrogen in the atmosphere based on the visibility limit of hydrogen at a given effective temperature. Since the hydrogen abundance cannot be measured directly in most metal-polluted white dwarfs, there exists no empirical relationship from which it can be estimated. We add a fixed hydrogen abundance of $\log\mathrm{H/He}=-3$, thus bringing the measured DZ masses closer to 0.6\msun, but also inducing a temperature-dependant offset (see the mass-temperature diagram in Figure \ref{fig:Mteff}). 

\subsection{Adopted Parameters}\label{sec:catalogue}
The effective temperature, mass, surface gravity, and chemical composition (see below) for all white dwarfs in our sample are all available via the online catalogue accompanying this paper. See Table \ref{tab:cat} for the full list of columns and their description. For objects that did not converge during the fitting procedure, the physical parameters are set to $-$999. Also included are the probabilities of belonging to each class ($P_\mathrm{class}$), the synthetic SDSS magnitudes and their flux errors as well as the $u$ band correction ($\Delta u$, see Section \ref{sec:photo}). The $u$ band correction bins shown in Figure \ref{fig:ucorr} are also available as Supplementary Material.

We briefly summarize the atmosphere assumptions for each spectral type as well as important points about the fitting procedure, if necessary. For DA stars, we fit the photometry assuming a pure hydrogen composition ($\mathrm{He/H}=0$), whereas for DB and DO stars, we assume a pure helium composition ($\mathrm{H/He}=0$). For DC white dwarfs, we initially fit all objects with a pure helium atmosphere, and fit them a second time based on the effective temperature, assuming a pure helium atmosphere above 11,000 K, a mixed atmosphere ($\log \mathrm{H/He}=-5$) between 11,000 K and 5500 K, and a pure hydrogen atmosphere below 5500 K. For DQ stars, we initially fit all objects with a helium-dominated atmosphere and a fixed carbon abundance ($\log \mathrm{C/He}=-5$), estimate a new carbon abundance based on empirical $\Te-\log \mathrm{C/He}$ relations and fit the photometry a second time using this new abundance. Finally, for DZ stars, we assume a helium-dominated atmosphere with a fixed abundance of calcium ($\log \mathrm{Ca/He}=-9.5$; all other metals are scaled in chondritic proportion according to this abundance).

Most objects without physical parameters appear to be WD$+$MS binary systems for which the fitting procedure did not converge. Many hot white dwarfs, including DA and DO, also do not have physical parameters due to large errors in the $u$ band. More generally, the measured properties of hot white dwarfs (\gta 40,000 K) should be used with extreme caution, if at all.

\input{tables/catalogue}

\section{Selected Results}\label{sec:disc}

In this section, we look at the global properties of our GSPC-WD sample, beginning with the various mass distributions. 

\subsection{Mass Distributions}\label{sec:mdist}

\begin{figure*}[ht]
    \centering
    \includegraphics{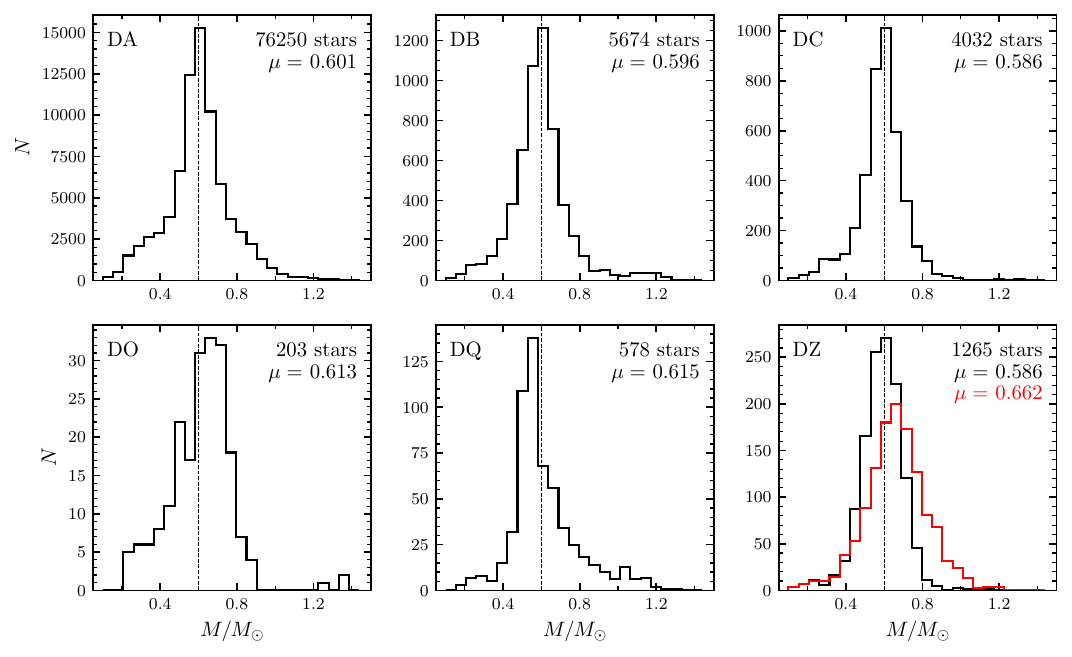}
    \caption{Cumulative mass distributions for the six spectral types in the GSPC-WD sample. The total number of stars in each histogram and the mean mass are shown at the top right corner. For DZ stars, the red and black histograms represent atmospheres with no hydrogen and a hydrogen abundance of $\log \rm{H/He}=-3$, respectively. The dashed line indicates the fiducial mean mass of 0.6\msun.}
    \label{fig:Mhist}
\end{figure*}

We show in Figure \ref{fig:Mhist} the cumulative mass distributions for our entire sample, $N$ versus $M$, for all spectral types (DA, DB, DC, DO, DQ, DZ), regardless of their effective temperature. In each panel we provide the number of stars as well as the mean mass $\mu$ of each subsample. Remarkably, all mass distributions are relatively narrow and peak near $\sim$0.6\msun, the fiducial mean mass for white dwarfs, indicated by a dashed line in each panel. Hence our overall classification scheme and fitting procedure seem to have properly captured the global properties of all spectral types. 

For the DA stars, there is a significant excess of low-mass and high-mass objects compared to other spectral types, consistent with the fact that low-mass white dwarfs, which are most likely unresolved double degenerate binaries, as well as high-mass white dwarfs, are mostly of the DA spectral type (see, e.g., Figure 15 of \citealt{caron2023} and Figure \ref{fig:Mteff_cool} below). The mass distributions of DA, DB, and DC stars all have a median mass close to 0.6\msun. Those of the DO, DQ, and DZ stars show a more complex behavior. The DO stars, although in relatively small number in our sample, show an extended tail towards low masses, and the median mass is shifted above 0.6\msun. This is a simple consequence of our difficulty with obtaining reliable physical parameters for these hot DO stars, as discussed above, a problem also apparent in the strong $M$ versus $\Te$ correlation observed in Figure \ref{fig:Mteff}. 

As expected, the median of the mass distribution for DQ stars is lower by about $\sim$0.05\msun\ compared to the median for other spectral types, a result also obtained earlier by
\citet[][see their Figure 13]{Coutu2019} and \citet[][see their Figure 19]{caron2023}. As discussed in Caron et al., there
are at least two possible explanations for this lower mean mass, one
involving problems with the physics of DQ model atmospheres
\citep{Coutu2019}, and another one recently proposed by
\citet{bedard2022}, who suggested that carbon -- and hence the DQ
phenomenon -- is preferentially detected in lower mass white dwarfs.

The mass distribution for DZ white dwarfs in Figure \ref{fig:Mhist} is shown for both an atmospheric composition assuming no hydrogen, as well as a trace of hydrogen of $\log \rm{H/He}=-3$. The effect of adding a trace of hydrogen in the analysis of DZ stars using the photometric technique is to decrease the stellar masses significantly, as first noted by \citet{Dufour2007}. In this case, the addition of free electrons from hydrogen changes the helium free-free opacity, resulting in lower photometric temperatures, and thus larger stellar radii and smaller masses. We thus assume a trace of hydrogen for DZ stars in the remainder of our analysis (and in our catalogue as well). Note that our photometric analysis of DC stars also include a trace of hydrogen, otherwise the peak of the mass distribution would be shifted towards higher masses as well.

Of more significant interest is the mass distribution of the various spectral types as a function of effective temperature. Here we focus our attention to the cool end ($\Te<10,000$ K) of the white dwarf sequence, and we also restrict our sample to a distance of 100 pc in order to reduce the number of objects in our plots, but also to compare our results directly with those of \citet{caron2023}, who restricted their analysis to the same distance. The $M$ versus $\Te$ distribution for this subsample is displayed in Figure \ref{fig:Mteff_cool} where we split the DA (top panel) and non-DA (bottom panel) stars for clarity. This figure can be compared directly with Figure 15 of \citet{caron2023}, who analyzed a significantly smaller sample of 2880 spectroscopically confirmed white dwarfs drawn from the MWDD, compared to the 12,569 objects in our 100 pc sample below 10,000 K. A note of caution here, however. Given that our classification scheme is strictly based not on spectral lines but on photometric information as well, the DA stars in the upper panel of Figure \ref{fig:Mteff_cool} also include at the end of the cooling sequence non-DA stars that are better fitted with pure H atmospheres (see also \citealt{caron2023}). 

\begin{figure*}[ht]
    \centering
    \includegraphics{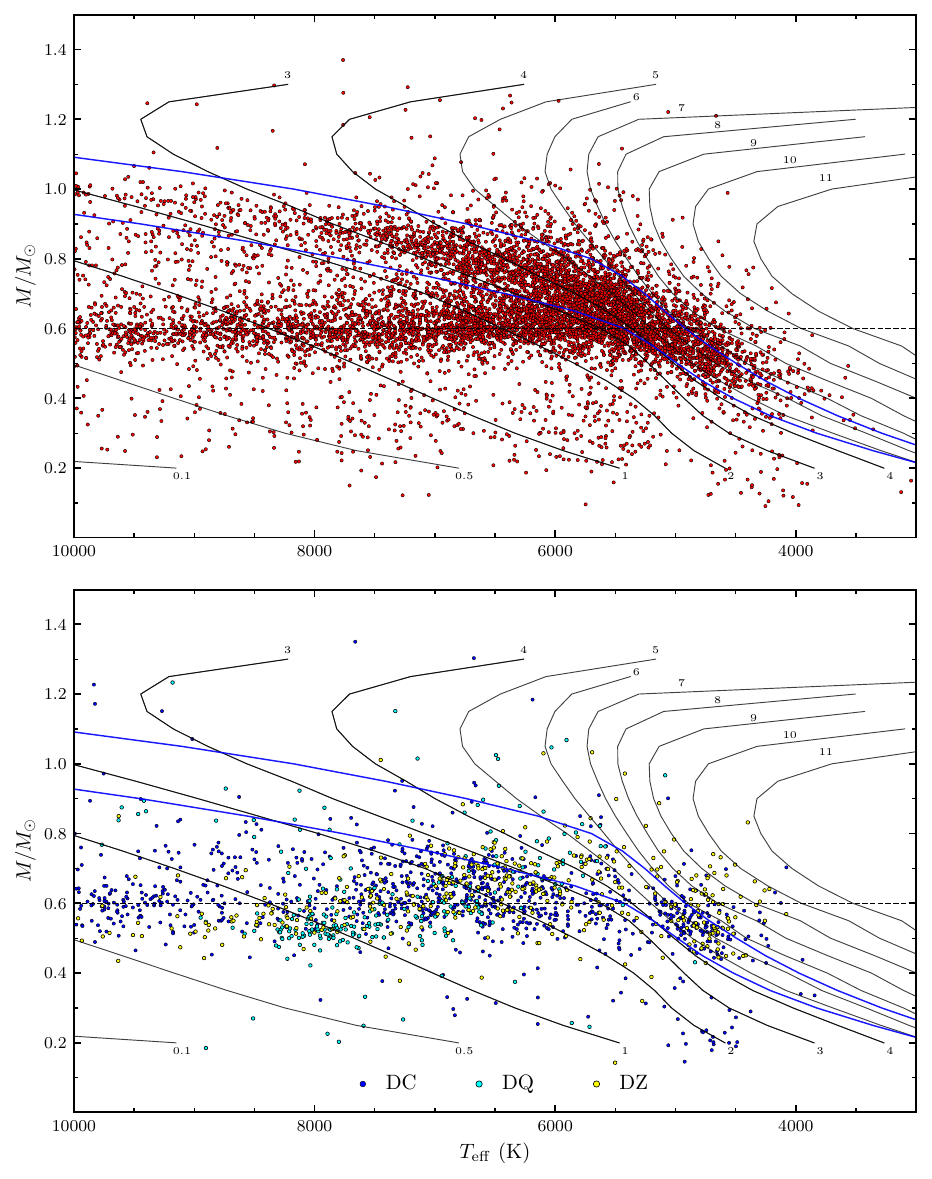}
    \caption{Mass-temperature diagrams of the GSPC cool white dwarfs within 100 pc of the Sun. The top panel shows the diagram for DA stars, and the bottom panel shows the non-DA stars. Also shown as solid black curves are theoretical isochrones, labeled in units of Gyr, obtained from cooling sequences with C/O-core compositions, $q(\rm{He})=10^{-2}$, and $q(\rm{H})=10^{-4}$. The lower blue solid curve indicates the onset of crystallization at the  centre of evolving models, and the upper one indicates the locations where 80\% of the total mass has solidified. The dashed line indicates the fiducial mean mass of 0.6\msun.}
    \label{fig:Mteff_cool}
\end{figure*}

The most striking feature in the mass distribution for DA stars is the crystallization sequence, which is contained between the two blue solid curves in Figure \ref{fig:Mteff_cool}, where the lower blue solid curve indicates the onset of crystallization at the  centre of evolving models, while the upper curve indicates the locations where 80\% of the total mass has solidified. This crystallization sequence evolves towards lower masses at lower $\Te$ values, and eventually merges with the other evolving DA stars with normal masses. Note that non-DA stars can also be found within this crystallization sequence, but in significantly smaller number.

As previously discussed above, low-mass ($M<0.5$\msun) white dwarfs are mostly DA stars, although low-mass non-DA stars exist as well. \citet{caron2023} argued, based on their more refined spectro-photometric analysis, that these low-mass non-DA white dwarfs probably have hydrogen atmospheres, which would suggest that common-envelope evolution most likely produces white dwarf remnants that retain thick H layers.

For $\Te<5500$ K, it was assumed that most DC white dwarfs have pure hydrogen atmospheres, following the conclusions of \citet{caron2023}; we remind the reader that a fraction of these cool stars are classified as `DA' in Figure \ref{fig:Mteff_cool}. The masses for these objects obtained under the assumption of helium atmospheres are way too low from an astrophysical point of view (see Caron et al.~for a more detailed discussion). We can see that at the very cool end of the DA sequence, the masses decrease slightly, a problem that has been attributed to inaccuracies in the calculations of opacity sources, either the red wing of L$\alpha$, the H${^-}$ bound-free opacity, the collision-induced opacity from molecular hydrogen, or several of the above.

The mass distributions for the DQ and DZ stars in the bottom panel of Figure \ref{fig:Mteff_cool} need to be interpreted with caution since for these objects, we adopted an empirical relation to fix the carbon abundance in DQ stars, and a fixed abundance of hydrogen in all DZ stars, while the exact abundance of each of these trace elements should in principle be adjusted individually for each object. For instance, the trend observed here, where the masses of both spectral types are larger at lower temperatures, is most likely the result of our assumptions.

Finally, we note that one spectral type that has been omitted from our analysis are the so-called ultracool white dwarfs, or more accurately described as IR-faint white dwarfs, which are characterized by a strong infrared flux deficiency resulting from collision-induced absorption by molecular hydrogen (see the detailed analysis by \citealt{Bergeron2022}), but these are difficult to identify in our sample due to the lack of infrared photometry in our analysis.

While the analysis of \citet{caron2023} relied on a more detailed and tailored analyses of individual objects of the 100 pc sample from the MWDD, the authors concluded the they had reached the limit of human capacity to analyse individually each object in their large sample of nearly 3000 white dwarfs. They also concluded that better techniques for handling bigger data sets involving machine-learning algorithms would eventually become necessary. Given that the results presented in this section compare favorably well with those of Caron et al.~in terms of the mass distributions, we believe that such a goal has now nearly been achieved.

\subsection{Spectral evolution}
The atmospheric composition of a white dwarf star can change as it evolves along the cooling sequence, a phenomenon referred to as the spectral evolution. Changes in atmospheric composition provide evidence for transport mechanisms competing with gravitational settling in determining the chemical appearance of the stars as they cool down. Here, we study the spectral evolution of the GSPC-WD sample by looking at the ratio of non-DA to the total number of stars as a function of effective temperature. We obtain a ratio for bins of 1000 K by summing the number of objects weighted by $1-P_\mathrm{DA}$ and by dividing by the unweighted total. The error bars are estimated using the Clopper-Pearson interval method \citep{clopperpearson}. We restrict our analysis to stars under 30,000 K due to the $u$ band calibration issues, described in the previous sections, affecting temperature measurements. We also exclude stars with masses below 0.45\msun\ as they mostly include unresolved binary systems \citep{Bergeron2019}. This leaves a total of 14,679 objects. A completeness estimation of the Gaia white dwarfs with XP spectra within 100 pc of the Sun has already been performed by \cite{Jimenez-Esteban2023}, who estimated that about 18,800 white dwarfs should be found within this distance based on the space-density obtained by the 20-pc sample of \cite{Hollands2018}. Under this assumption, our sample is $\sim$78\% complete. We can also make a rough estimate of our selection completeness by comparing the number of objects to the white dwarf candidate catalogue of \cite{GentileFusillo2021}, who estimated their overall completeness to be between 67\% and 93\%. Assuming that the 16,675 objects with $P_{\rm WD}>0.75$ within 100 pc of the Sun in the \cite{GentileFusillo2021} catalogue represent a 93\% volume-complete sample, which is likely appropriate at the said distance, our own selection would be $\sim$82\% complete. Based on the two different completeness determinations above, we estimate our 14,679 stars to represent a $\sim$80\% volume-complete sample.

\begin{figure}[ht]
    \centering
    \includegraphics{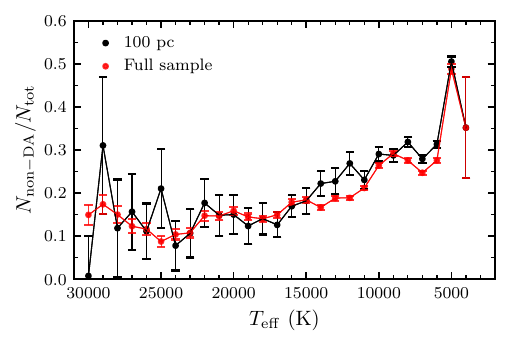}
    \caption{Spectral evolution curves of the GSPC-WD sample for objects within 100 pc of the Sun (black line) and the entire sample (red line). The non-DA ratio is calculated by summing the number of non-DA white dwarfs in each temperature bin, weighted according to their probability of being a DA, divided by the total number of stars within the bin.}
    \label{fig:specevol}
\end{figure}

The spectral evolution of the GSPC-WD sample is shown in Figure \ref{fig:specevol} for objects within 100 pc (black curve) and the full sample (red curve). The latter is only shown as a reference, since it suffers from important selection biases that are not corrected for here. The discussion below focuses on the 100 pc sample. We first notice that both curves appear to systematically overestimate the ratio of non-DA by $\sim5\%$ when compared to previous studies \citep[see Figure 7 of][]{Torres2023}. This offset can easily be eliminated by summing the number of objects above or below certain classification thresholds rather than weighting them by their classification probability. However, the former approach would result in the loss of crucial statistical nuance at lower temperatures, where the classification boundaries are not well defined. For consistency, we stick to the weighted sum approach and keep in mind that the ratio values might be slightly overestimated.

At high temperatures, we find large fluctuations of the ratio of non-DA. This is likely a small number statistics effect, as very few helium-atmosphere objects are found above 25,000 K \citep{GBB2019}. This can also be seen on the mass-temperature diagrams of Figure \ref{fig:Mteff}, where only a few DBs reside in the 30,000-25,000 K temperature range. Increasing the volume limit to include more objects stabilizes the number to 15\%, which connects nicely with the high-temperature spectral evolution sequence in \cite{bedard2020} when accounting for the systematic offset mentioned above. \cite{bedard2020} proposed that this gradual decrease can be explained using the float-up model, where a broad range of residual hydrogen diffuses to the surface and turns most helium-rich stars into DA before they reach $\Te\sim30,000$ K. 

A bump of non-DA white dwarfs between 30,000 K and 25,000 K followed by a deficit between 25,000 K and 22,000 K was reported by \cite{Jimenez-Esteban2023} and \cite{Torres2023} who analyzed the 100 pc and 400 pc volume-limited samples, respectively, of the GSPC-WD catalogue. According to their analysis, the ratio of non-DA starts increasing around 30,000 K, reaching its peak at 27,500 K with an increase of $\sim$5\% in non-DA stars, and goes back down to its previous ratio at 25,000 K. We do not find any evidence for such a bump nor a deficit in our 100 pc sample. Instead, we find the non-DA ratio to be statistically constant between 29,000 K and 18,000 K with a ratio somewhere between 10\% and 20\%. While the spectral evolution of the full GSPC-WD in Figure \ref{fig:specevol} may be suggestive of the aforementioned bump and deficit, they do not exhibit precisely the same characteristics as the features identified by \cite{Jimenez-Esteban2023} and \cite{Torres2023}. Furthermore, it is essential to consider that the full GSPC-WD sample is not yet volume-complete, as demonstrated in Figure \ref{fig:completeness}, and we have yet to account for potential selection biases. Notably, the color-dependent calibration issues discussed in Section \ref{sec:atmo}, which impact the physical parameters of white dwarfs beginning around 30,000 K (as evident in the DA and DO mass-temperature diagrams in Figure \ref{fig:Mteff}), were not addressed by the authors. The implications of these calibration issues on the spectral evolution beyond 30,000 K remain uncertain; however, it is plausible that they could sufficiently shift the temperature values, potentially leading to a subtle perturbation resembling a small bump. Adding to this, it is worth noting that the classification between DA and non-DA in \cite{Jimenez-Esteban2023} and \cite{Torres2023} relies primarily on the optimal photometric fit between DA and non-DA spectra. This classification method has been shown in the past to have diminishing accuracy as effective temperature exceeds approximately 25,000 K \citep[see Figure 2 in][]{Bergeron2019}. Consequently, even a minor calibration offset has the potential to introduce significant classification inaccuracies.

Moving on to cooler effective temperatures, the ratio of non-DA steadily increases between 18,000 K and 8000 K, consistent with previous studies \cite{Ourique2020}, \cite{Cunningham2020}, \cite{Jimenez-Esteban2023}, and \cite{Torres2023}. This is indeed what is expected from the convective dilution and convective mixing processes taking place within that temperature range \citep{Rolland2018, GBB2019, Cunningham2020}, causing a gradual transformation of DA into non-DA stars. We note a peculiar dent at the 7000 K bin. While small, it appears statistically significant and goes against the upward trend observed in low-temperature spectral evolution studies \citep{Blouin2019, McCleery2020}. This feature is greatly accentuated when we calculate the non-DA ratio using simple sums rather than weighted sums, causing the non-DA ratio to rapidly drop between 10,000 K and 6000 K before sharply going back up at 5000 K. Due to its sensitivity on how the non-DA ratio is calculated, we are cautious to affirm whether this dent is physically meaningful. It does, however, persist in the full GSCP-WD sample spectral evolution in Figure \ref{fig:specevol}. A possible explanation for this dent could simply be that XP spectra become increasingly difficult to classify at low temperatures. We looked at the prediction confidence of each class as a function of temperature for all objects, and found that the classifiers indeed become increasingly confused at around 9000 K. Another factor that could contribute to this dent are that non-DA may be turning into hydrogen-rich DC stars at exactly this temperature \citep{Kowalski2006, caron2023}. Of course, one should also consider the fact that low-temperature objects are also fainter, and that the measured temperature is more uncertain. Furthermore, it is well known that the physics of very cool white dwarfs is still missing important pieces below 6000 K \citep{Saumon2022}, and any physical parameter measured for the coolest white dwarfs should be interpreted with utmost care.

Finally, at the lowest temperatures, we find the ratio of non-DA to increase sharply to $\sim$50\% at 5000 K, then dropping back down to $\sim$35\% at 4000 K. This behavior is consistent with the expectation that white dwarfs tend to become increasingly helium-rich as they cool down \citep{Blouin2019, McCleery2020} and then have their atmospheres turn into hydrogen-rich, albeit we find the latter transition to happen at a lower temperature than what was found by \cite{caron2023}. We hypothesize that the DA classifier gradually starts to rely on the slope of the stellar continuum rather than absorption lines as it tries to classify fainter objects, but only realizes to do so around 4000 K, where absorption lines are certain to be absent. As a matter of fact, we find that if the non-DA ratio is calculated by simply summing all the stars classified as non-DA, the transition from helium-rich to hydrogen-rich atmospheres begins at the expected temperature range between 6000 and 5000 K \cite{caron2023}. As noted earlier, white dwarfs at low temperatures become increasingly difficult to interpret and the manner in which the ratio of non-DA is calculated has significant impact on the shape of the spectral evolution. 

To summarize, we studied the observed spectral evolution of the 100 pc GSPC-WD sample for stars with masses above 0.45\msun\ and effective temperatures between 30,000 K and 5000 K. Albeit our non-DA ratios display a systematic overestimation of about 5\% compared to previous results in the literature, which can be explained by the use of weighted sums, the global trends remain the same. Of particular note are the lack of a non-DA bump and deficit between 30,000 K and 22,000 K, as well as a small dent at 7000 K.

\subsection{White Dwarf Luminosity Function}

We present here the 100 pc observed white dwarf luminosity function (WDLF) of the GSPC-WD, meaning no corrections are applied due to the incompleteness of the survey. For the same reasons outlined in the previous section, we restrict our selection to stars with measured masses above 0.45\msun\ and temperatures below 30,000 K and above 5000 K, resulting in an approximately 80\% volume-complete sample.

The WDLF is a measure of the number of stars per pc$^3$ per unit of bolometric magnitude, which we obtain using the luminosity ($L/\lsun$) derived from the photometric results of white dwarfs within 100 pc provided in Table \ref{tab:cat}. The bolometric magnitudes are calculated using the relation $M_\mathrm{bol} = - 2.5 \log L/\lsun + \mbol^\odot$, where $\mbol^\odot=4.75$ is the bolometric magnitude of the Sun. Each object in the sample is then simply added to the appropriate bolometric magnitude bin, and the overall results are divided by the volume defined by a 100 pc sphere.

\begin{figure}[ht]
    \centering
    \includegraphics{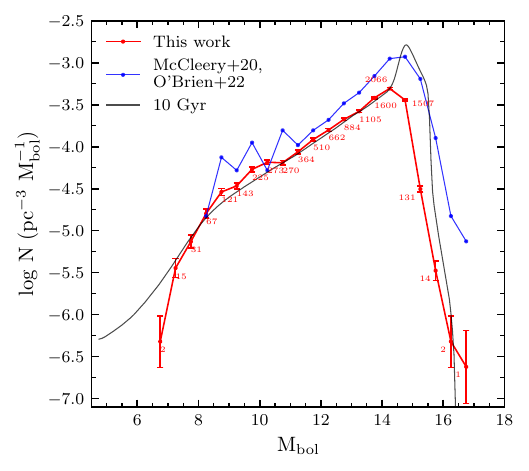}
    \caption{White dwarf luminosity functions of GSPC white dwarfs within 100 pc of the Sun (red line) and spectroscopically confirmed white dwarfs within 40 pc of the Sun \citep[blue line; ][]{McCleery2020, Obrien2023}. Error bars represent the Poisson statistics of each bolometric magnitude bin. Also shown for reference is the theoretical luminosity function from \cite{Fontaine2001} for a total age of 10 Gyr (black line).}
    \label{fig:WDLF}
\end{figure}

The luminosity function for the GSPC white dwarfs within 100 pc of the Sun is presented in Figure \ref{fig:WDLF}. Our results are also compared with the volume-complete spectroscopic survey of white dwarfs within 40 pc the Sun by \cite[][northern survey]{McCleery2020} and \cite[][southern survey]{Obrien2023}. We take the photometric effective temperature and surface gravity provided in the respective papers and convert them into luminosity and bolometric magnitude using the evolutionary models described in Section \ref{sec:atmo}. We also include, for reference only, the theoretical luminosity function from \cite{Fontaine2001} and \cite{Limoges2015} for a total age of 10 Gyr, normalized to our own observational results between $\mbol=14.0-14.5$. Briefly, the theoretical luminosity functions were obtained using a constant star formation rate, a classic Salpeter initial mass function ($\phi=M^{-2.35}$), an initial-to-final mass relation given by $M_\mathrm{WD} = 0.4e^{0.125M}$, a main sequence lifetime law given by $t_{\rm MS}=10M^{-2.5}$ Gyr, where $M$ and $M_{\rm WD}$ are in solar units.

Starting with the hot end of the observed luminosity function, we find a sharp increase of the space density between bolometric magnitudes 7 and 8, which may be partially caused by our effective temperature cut. We find that including hotter objects up to 40,000 K pushes the space densities closer to what is predicted by the the theoretical luminosity function of \cite{Fontaine2001}, but a statistically significant dent remains nonetheless. We suspect this feature may be explained by a large 100 pc volume, as we move beyond the thin disk and start reaching more deeply into the thick disk, where the star population is older and the space density diminishes. This increase between 7 and 8 mag was also found in the spectroscopic analysis of white dwarfs found in the Kiso survey by \cite{Limoges2010} and in the deep proper motion survey \citep[][see their Figure 12]{Munn2017}, where both studies combined both thin and thick disk objects in their luminosity functions. In contrast, studies that focus on the thin disk find a much smoother increase of white dwarf space density \citep{Harris2006, DeGennaro2008, Krzesinski2009}. Further lending support to this idea is the more recent study by \cite{Kilic2017}, who have shown that distinctly considering the thin and thick disk populations and combining them assuming a 20\% thin/thick ratio could better reproduce the brightest part of the observed luminosity function in \cite{Munn2017}. One can visualize the impact of considering the two disks separately with the theoretical luminosity function in Figure \ref{fig:WDLF}, which does not make the distinction between the two, and predicts a smoother increase that is appropriate for the think disk population.

Moving on to the middle section of the white dwarf luminosity function, theoretical models usually predict a monotonic rise between 8 and 15 mag. As seen in Figure \ref{fig:WDLF}, our observed luminosity function perfectly follows the theoretical one, apart from the small bump around $\mbol\sim10$. This bump was first noticed almost two decades ago using SDSS data \citep{Harris2006}, and has since then also been found in numerous other studies, including ours and the latest spectroscopic sample of white dwarfs within 40 pc of the Sun (blue line in Figure \ref{fig:WDLF}). The currently accepted explanation was originally suggested by \cite{Limoges2015}, who proposed that the bump can be explained by enhanced star formation around 300 Myr ago. The hypothesis was further explored by \cite{Torres2016}, who revisited the analysis of the 40 pc sample with a population synthesis code, as well as newer initial-to-mass relation and cooling tracks. They also explained the bump around $\mbol=10$ mag as a burst of star formation 600 Myr ago, however, their best-fit model significantly over-predicts the number of white dwarfs near the maximum of the luminosity function. \cite{Torres2016} explained this discrepancy with an initial–final mass relation that has a slope 30\% larger than the observed relation for stars more massive than 4\msun\ from \cite{Catalan2008}. As pointed out by \cite{Kilic2017}, there is no evidence for such a steep initial–final mass relation and this explanation is unlikely. Instead, \cite{Kilic2017} proposed the contribution of thick disk white dwarfs to the faint end of the luminosity function as a possible explanation for the overabundance of white dwarfs near the maximum of the luminosity function.

At the faint end of our observed luminosity function, we find the peak at $\mbol=15$ mag, followed by the so-called dropoff \citep{Fontaine2001}. This region is the most important when it comes to measuring the age of the galactic components and is the most sensitive to selection biases, white dwarf physics, and assumptions behind the theoretical luminosity function \citep[see][for a review on the subject]{GBO2016}. An in-depth analysis of the faint end of the luminosity function is beyond the scope of this paper and will be done in future work. Here, we limit our analysis to the global features and compare them with the volume-complete 40 pc spectroscopic sample and theoretical luminosity function in Figure \ref{fig:WDLF}. 

The most obvious difference is, perhaps, the less pronounced peak of our observed luminosity function. It is about 17\% smaller in maximum amplitude than the 40 pc sample and definitely does not follow the trend predicted by \cite{Fontaine2001}. The difference with the 40 pc sample can likely be explained due to incompleteness and selection biases, which become particularly important for fainter stars. As we go down in effective temperature, the classifiers become increasingly confused and the proportion of objects that fall below the classification confidence threshold increases, thus reducing the space density we find. On the astrometric side, the combination of thin and thick disks can significantly alter the shape of the white dwarf luminosity function near its peak \citep{Kilic2017} and may produce what we observe here. It is unsurprising, however, that our results strongly differ from the theoretical predictions of \cite{Fontaine2001} near the peak, as their physics and population synthesis were rudimentary and have since been updated and predict a much flatter peak, closer to what we find \cite[see ][and references therein]{Tononi2019}. Another detail to consider is our fixed model atmosphere assumptions that affect the measured physical parameters. For example, the fixed metal and hydrogen abundances in our DZ fits cause small systematic effects (see Sections \ref{sec:atmo} and \ref{sec:mdist}). A larger average mass would imply smaller radii and lower luminosities, thus shifting the stars to fainter magnitude bins. 

Finally, one can also see from Figure \ref{fig:WDLF} that the dropoff of our observed luminosity function happens about 0.5 mag earlier than the 40 pc sample. This can likely be attributed to the simple fact that our WDLF includes a much larger number of objects from the thick disk, which is well-known to contain an older white dwarf population \citep{GBO2016, Kilic2017}.

To conclude, the GSPC-WD catalogue offers a nearly volume-complete sample within 100 pc of the Sun with considerable potential for the study of the WDLF. In particular, disentangling the different galactic components and measuring their relative contribution to the total white dwarf luminosity function would be very interesting given the results of \cite{Kilic2017}. The faint end of the luminosity function would also make for an interesting in-depth analysis, as the GSPC-WD catalogue offers the largest sample of cool white dwarfs with spectroscopic classifications, a step forward compared to previous studies that typically assume a pure hydrogen or pure helium atmosphere for white dwarf candidates. A more detailed analysis of the GSPC WDLF is currently under way and should be published soon.

\section{Conclusion}\label{sec:conc}
In this paper, we have spectroscopically classified $\sim$100,000 white dwarf XP spectra as one of six possible types (DA, DB, DC, DO, DQ, DZ) and measured their physical parameters using synthetic SDSS photometry, nearly tripling the number of white dwarfs with spectroscopic classification. We summarize the major results from this work below:

\begin{itemize}
    \item We have demonstrated that XP spectra have sufficient resolution to achieve reliable classification across the six spectroscopic types listed above, for most stars. We have validated the classifications using the Gaia H-R diagram and by recovering many cornerstone features in the physical parameter distributions, the spectral evolution and the luminosity function of the GSPC white dwarfs.
    \item We have measured the physical parameters of the white dwarfs using class-appropriate model atmospheres. We recover the expected mass and temperature distributions obtained by higher-resolution spectroscopic studies applied to smaller numbers of objects, extending them to include a much larger sample of stars. This is a significant step beyond simply assuming pure hydrogen or mixed atmospheres for the Gaia white dwarf candidates.
    \item The blue region of the XP spectra currently suffers from color-dependent calibration issues, causing large systematic errors in the measurement of physical parameters of the hottest white dwarfs. We calculated a correction to the synthetic $u$ band to alleviate this issue. However, despite this correction, we advise caution when interpreting results above $\Te=40,000$ K.
    \item The spectral evolution of GSPC white dwarfs within 100 pc of the Sun was studied through the ratio of non-DA as a function of effective temperature. We found that the spectral evolution closely resembles what was found in previous studies. Interestingly, we also found evidence of the expected helium-to-hydrogen atmosphere transition at very low temperatures.
    \item We have carried out an initial study of the observed GSPC white dwarf luminosity function and found evidence supporting the idea that distinctly considering the thin and thick disks could help explain some of its features. Given the high level of completeness of the GSPC-WD 100 pc sample and our atmosphere-appropriate measurement of the physical parameters of the white dwarfs, the white dwarf luminosity function we have obtained is ripe for a new and deeper volume-limited analysis, which is currently underway.
\end{itemize}

The DR3 release from the Gaia mission is already providing us with exciting opportunities to perform statistical studies of large samples of white dwarf stars. The DR4 release, planned for late 2024 or early 2025, will provide an even larger number of XP spectra and will increase the quality of the current sample by including an extended data collection period delivering consequent longer exposures. It will be in the interests of the community to prepare for this incoming data set by preparing tools such as improved classifiers, trained on synthetic data, and methods to measure the physical parameters directly from the XP spectra coefficients rather than byproducts, such as synthetic photometry or sampled spectra.

\begin{acknowledgements}
The authors thank Antoine B\'edard for useful discussions on the topics of spectral evolution and hot white dwarfs. This work presents results from the European Space Agency (ESA) space mission Gaia. Gaia data are being processed by the Gaia Data Processing and Analysis Consortium (DPAC). Funding for the DPAC is provided by national institutions, in particular, the institutions participating in the Gaia MultiLateral Agreement (MLA). The Gaia mission website is https://www.cosmos.esa.int/gaia. The Gaia archive website is https://archives.esac.esa.int/gaia. This work is supported in part by the United Kingdom Space Agency (Grants: ST/K000578/1, ST/N000978/1, ST/S001123/1, ST/W002809/1, ST/X001687/1), NSERC Canada and by the Fund FRQ-NT (Qu\'ebec).

\end{acknowledgements}

\section*{Data Availability}
The data that support the findings of this study are openly available. The spectroscopic and astrometric data can be accessed from the official Gaia archive servers\footnote{\url{https://gea.esac.esa.int/archive}}. A full description of the catalogue produced by this paper can be found at Table \ref{tab:cat} and can be downloaded online\footnote{www.astro.umontreal.ca/{\textasciitilde}ovincent/catalogues}, along with the $u$ band correction bins shown in Figure \ref{fig:ucorr}. The results of this paper will also be made available both on the MWDD website\footnote{montrealwhitedwarfdatabase.org} and the publisher website.

\bibliographystyle{aa} 
\bibliography{bibfile_for_XP_paper.bib}

\begin{appendix}
\section{Classification algorithm details and data processing}\label{appendix:classifier}
This appendix outlines the necessary details in order to reproduce our results using the classification approach and data described in Section \ref{sec:classifier}. First, the 110 XP coefficients are normalized to $G=15$ following the recommendations in \cite{Andrae2023}. This is done by dividing all the coefficients of each star by $10^{15-G} / 2.5$. For each binary classifier, we then weight the labels according to the number of objects in each class, e.g. DA and non-DA for the DA classifier, DB vs non-DB for the DB classifier, and so on. The weights of each class ($w_{\rm class}$) are calculated using the sample formulae $w_{\rm class} = 1/2N_{\rm class}$. We have found that weights for very imbalanced classes, such as DO stars, become very large and cause performance issues. We thus bring large weights ($w_{\rm class} > 10$) back to a more reasonable value by dividing them by 10, resulting in better performance.

As for the classifier hyperparameters, we use the same setup for every classifier and keep most of the default options set in the \texttt{GradientBoostingClassifier} object as of version 1.1.2 of Scikit-learn. The only changes in parameters were the number of estimators (\texttt{n\_estimators=225}), the learning rate (\texttt{learning\_rate=0.1}) and the maximum tree depth (\texttt{max\_depth=3}).
\end{appendix}

\end{document}

%% file: tables/classifier2.tex
\begin{table}
\centering
\caption{Training examples and average precision and recall test scores of the cross-validation and top-5 classifier ensembles at a 0.6 threshold.}
\label{tab:classifier}
\begin{tabular}{lccccc}
\hline
Class & $N$ & $P_\mathrm{c-v}$ & $R_\mathrm{c-v}$ & $P_\mathrm{top5}$ & $R_\mathrm{top5}$ \\
\hline \hline
DA & 10530 & 0.97 & 0.98 & 0.98 & 0.98 \\
DB & 1323 & 0.96 & 0.93 & 0.96 & 0.95 \\
DC & 884 & 0.64 & 0.68 & 0.66 & 0.72 \\
DO & 46 & 0.76 & 0.76 & 0.95 & 0.90 \\
DQ & 350 & 0.57 & 0.88 & 0.60 & 0.89 \\
DZ & 412 & 0.60 & 0.85 & 0.62 & 0.86 \\
\hline
\end{tabular}
\end{table}

%% file: tables/GSPC-WD.tex
\begin{table}
\centering
\caption{High-confidence white dwarfs per class in the GSPC-WD sample and in the Gaia-SDSS catalogue.}
\label{tab:GSPCWD}
\begin{tabular}{llll}
\hline
\multicolumn{1}{l}{Class} & \multicolumn{1}{c}{$N$} & \multicolumn{1}{c}{$N_\textrm{SDSS}$} & \multicolumn{1}{c}{$N_\textrm{new}$}\\
\hline\hline
DA & 77330 & 20891 & 69403 \\
DB & 5688 & 1650 & 4797 \\
DC & 4082 & 1930 & 3346 \\
DO & 215 & 69 & 183 \\
DQ & 601 & 406 & 360 \\
DZ & 1272 & 896 & 958 \\
\hline
\end{tabular}
\end{table}

%% file: tables/catalogue.tex
\begin{table}
 \caption{Columns and descriptions of our online catalogue for the GSPC-WD sample.}
 \label{tab:cat}
 \centering
 \begin{tabular}{lll}
    \hline
    Column Header & Description \\
    \hline
    \texttt{source\_id} & Gaia DR3 source identification number\\
    \texttt{spectype} & Primary spectroscopic type (Sec. \ref{sec:classifier})\\
    \texttt{teff} & Effective temperature (K)\\
    \texttt{teff\_err} & Error on the effective temperature (K)\\
    \texttt{logg} & Logarithm of surface gravity\\
    \texttt{logg\_err} & Error on the surface gravity\\
    \texttt{M} & Mass (\msun) \\
    \texttt{M\_err} & Error the mass (\msun)\\
    \texttt{logL} & Log of the luminosity ($\lsun$) \\
    \texttt{logL\_err} & Error on the Log luminosity ($\lsun$) \\
    \texttt{logCHe} & Carbon abundance for DQ stars\\
    \texttt{comp} & Model atmosphere composition (Sec. \ref{sec:atmo})\\
    \texttt{mag\_u} & Synthetic SDSS u magnitude\\
    \texttt{mag\_g} & Synthetic SDSS g magnitude\\
    \texttt{mag\_r} & Synthetic SDSS r magnitude\\
    \texttt{mag\_i} & Synthetic SDSS i magnitude\\
    \texttt{mag\_z} & Synthetic SDSS z magnitude\\
    \texttt{flux\_error\_u} & Synthetic SDSS u flux error\\
    \texttt{flux\_error\_g} & Synthetic SDSS g flux error\\
    \texttt{flux\_error\_r} & Synthetic SDSS r flux error\\
    \texttt{flux\_error\_i} & Synthetic SDSS i flux error\\
    \texttt{flux\_error\_z} & Synthetic SDSS z flux error\\
    \texttt{P\_DA} & Probability of being a DA\\
    \texttt{P\_DB} & Probability of being a DB\\
    \texttt{P\_DC} & Probability of being a DC\\
    \texttt{P\_DO} & Probability of being a DO\\
    \texttt{P\_DQ} & Probability of being a DQ\\
    \texttt{P\_DZ} & Probability of being a DZ\\
    \texttt{u\_corr} & u band correction (Sec. \ref{sec:photo})\\
    \texttt{u\_corr\_err} & u band correction error (Sec. \ref{sec:photo})\\
  \hline
 \end{tabular}
\end{table}